\documentclass[aps,prb,reprint,twocolumn,superscriptaddress,longbibliography]{revtex4-2}

\usepackage[utf8]{inputenc}
\usepackage{bm}
\usepackage{graphicx}
\usepackage{amsmath,amsfonts,amssymb}
\usepackage[colorlinks=true,citecolor=blue,linkcolor=blue,urlcolor=blue]{hyperref}
\usepackage{xcolor}

\newcommand{\uAnm}{\,\mu\mathrm{A\,nm/V}^2}
\newcommand{\eV}{\,\mathrm{eV}}
\newcommand{\bk}{\mathbf{k}}
\newcommand{\bq}{\mathbf{q}}
\newcommand{\Rv}{\mathbf{R}}
\newcommand{\Rb}{\bar{\mathbf{R}}}
\newcommand{\sig}{\sigma^{y;yy}}
\newcommand{\svort}{\sigma_{\mathrm{vort}}}
\newcommand{\sberry}{\sigma_{\mathrm{Berry}}}
\newcommand{\sgrad}{\sigma_{\mathrm{grad}}}
\newcommand{\sharm}{\sigma_{\mathrm{harm}}}
\newcommand{\BZ}{\mathrm{BZ}}
\newcommand{\Kk}{\boldsymbol{\mathcal K}}
\newcommand{\Ll}{\boldsymbol{\mathcal L}}

\graphicspath{{figures/}{./}}

\begin{document}

\title{Exact vortex decomposition of the shift current}

\author{Mat\'ias Castro Schnaidt}
\affiliation{Departamento de F\'isica, Universidad T\'ecnica Federico Santa
Mar\'ia, Casilla 110-V, Valpara\'iso, Chile}
\author{Felipe P\'erez Riffo}
\affiliation{Departamento de F\'isica, Universidad T\'ecnica Federico Santa
Mar\'ia, Casilla 110-V, Valpara\'iso, Chile}

\author{Juan M. Florez}
\affiliation{Facultad de Ingenier\'ia y Arquitectura, Universidad Central de Chile, Santiago, Chile}

\author{Eric Su\'arez Morell}
\email{eric.suarez@usm.cl}
\affiliation{Departamento de F\'isica, Universidad T\'ecnica Federico Santa
Mar\'ia, Casilla 110-V, Valpara\'iso, Chile}

\date{\today}

\begin{abstract}
Does any integer survive in the shift current, and does the Chern number control its sign? We show that the shift vector is a singular field on the Brillouin torus. Its curl has two sources: the interband Berry curvature and quantised point charges at the zeros of the transition dipole. Inverting this relation gives an exact decomposition of the shift conductivity, valid at every frequency, into one sector linear in the integer charges and three continuous sectors. As a testing ground, we take a graphene-Haldane bilayer: a model that lets us sweep the interband Chern number through five values at fixed gap to isolate its role. We also test the formalism against the time-reversal-symmetric biased AB bilayer. Whenever a charge sits at the band edge, and trigonal warping is present, the integer sector alone accounts for the band-edge peak. The relevant integer is not the Chern number but the winding of that single charge. Its sign, combined with a valley-dependent profile, sets the sign of the photocurrent. In these bilayers, sign reversals across topological transitions follow this winding; the invariant changes with it only when the same gap closing does both. A band-edge charge needs no invariant. It only requires that the two band-edge sublattices not be connected by the nearest-neighbour hopping, and trigonal warping makes it visible.\end{abstract}

\maketitle

\section{Introduction}

The shift current is a second-order photocurrent fixed by the quantum geometry of the Bloch bands. Unlike the trace of the circular injection current of a Weyl node, which is quantised by its monopole charge~\cite{deJuan_2017, Ahn_2020}, the shift conductivity at a given frequency has no known quantisation. Quantised quantities do appear in the shift-current problem, as Berry phases of the shift vector along special lines~\cite{Alexandradinata_2024}, loop integrals of the shift vector~\cite{wangchang} or frequency integrals in multigap phases~\cite{Jankowski_2024}, but in the spectrum itself the shift vector enters weighted by the transition dipole and no integer survives the integral. Sign reversals of the shift current across topological transitions have been reported in several two-band models~\cite{Tan_2016, Sivianes_2023, LinHsu_2025}, but without a quantitative link between the invariant and the spectrum.

Here we make that link exact. The shift vector is a singular vector field on the Brillouin torus whose curl has two sources: the smooth interband Berry curvature and quantised point charges at the zeros of the transition dipole, the optical vortices, whose total winding is fixed by the Chern numbers~\cite{MeraOzawa_2022, Zhu_2024}. Inverting the curl as an electrostatics problem on the torus, a Helmholtz-Hodge decomposition splits the shift conductivity, at every frequency, into four sectors, one of which is linear in the integer charges. The topological content of the response is thereby not hidden but isolated.

The zeros of the transition dipole, which count the Chern number of a transition~\cite{MeraOzawa_2022}, have been identified as optical vortices that enhance photovoltaic currents~\cite{Zhu_2024, Sukhachov_2026}. In a previous work~\cite{noteAB} we derived the angular selection rule by which trigonal warping switches on the band-edge shift current of the AB bilayer, its reversal with the bias, and the growth of the band-edge response as $1/\delta$ in biased multilayers, with $\delta$ the layer bias. The present work is complementary: it gives the exact four-sector decomposition of the response, resolves the response charge by charge, finds it independent of the Chern number across five phases, and derives a per-valley sign rule and an analytic criterion for a charge at the zone corner.

Applied to a graphene bilayer with a Haldane layer, whose interband Chern number can be swept through five values at fixed resonant gap, and to the time-reversal-symmetric AB bilayer with a layer bias, the decomposition indicates that the integer sector carries the band-edge peak, with a small surplus that the continuous sectors remove; that at fixed warping the sign of the photocurrent is set by the winding of the single charge at the band edge, and that the Chern number itself orders neither the sign nor the magnitude. What matters is the local charge, which exists with or without an invariant, and the trigonal warping that switches on its spectral weight.

\section{Shift vector and optical vortices}

For a pair of bands $(v,c)$ the shift conductivity for current along $a$ and light linearly polarised along $b$ is~\cite{AversaSipe_1995, sipe2000, young2012}
\begin{equation}
\sigma^{a;bb}(\omega)=-\frac{2\pi e^3}{\hbar^2}\,g_s\!\int_{\BZ}\!\frac{d^2k}{(2\pi)^2}
\,\bigl|r^b_{cv}\bigr|^2\,R^{a;b}_{cv}\,\delta(\omega_{cv}-\omega),
\label{eq:sigma}
\end{equation}
with $g_s=2$ the spin degeneracy, $\hbar\omega_{cv}=E_c-E_v$ the transition energy and $\sigma$ defined through $J^a=\sigma^{a;bb}E_b(\omega)E_b(-\omega)$ for the complex field amplitudes. In two dimensions $\sigma^{a;bb}$ is a sheet conductivity, quoted in $\uAnm$; dividing it by an effective thickness gives a bulk value. With this definition $\sigma^{a;bb}$ is twice the conductivity defined through $J^a=2\sigma^{a;bb}E_b(\omega)E_b(-\omega)$, as in Ref.~\cite{Cook_2017}.
Two geometric quantities enter. The first,
\begin{equation}
    r^b_{cv} = \frac{v^{b}_{cv}}{i \omega_{cv}}, \qquad v^{b}_{cv} = \hbar^{-1}\langle u_c|\partial_{k_b}H|u_v\rangle ,
\end{equation}
is the interband transition dipole; the second is the shift vector
\begin{equation}
R^{a;b}_{cv}=\partial_{k_a}\arg r^b_{cv}+\bigl(A^a_{vv}-A^a_{cc}\bigr),
\label{eq:shiftvec}
\end{equation}
with $\mathbf A_{nn}=i\langle u_n|\nabla_{\bk}u_n\rangle$ the Berry connection. Each term of Eq.~\eqref{eq:shiftvec} is gauge dependent and the sum is gauge invariant; it is the real-space displacement of the charge centre in the vertical transition.

Fixing the polarisation $b$ and the pair $(v,c)$, the shift vector $\Rv(\bk)=(R^{x;b}_{cv},R^{y;b}_{cv})$ is a vector field on the Brillouin torus $T^2$. It is not smooth everywhere: it is singular at the zeros of the transition dipole, $|r^b_{cv}|^2=0$~\cite{Zhu_2024, Alexandradinata_2024, noteVortex}, around which the  dipole phase winds by an integer multiple of $2\pi$. We call these zeros the optical vortices and the integer $n_i$ the vortex charge, or winding, of the $i$-th zero. Cao, Wu, and Louie showed that the optical matrix elements of two-dimensional crystals, including graphene multilayers, carry winding numbers tied to band topology, and used them to derive excitonic selection rules~\cite{CaoWuLouie_2018}.

Taking the curl of Eq.~\eqref{eq:shiftvec}, the phase gradient contributes $2\pi n_i$ concentrated at each zero, since the phase of $r^b_{cv}$ winds by $2\pi n_i$ around it, and the two Berry connections contribute $\Omega_v-\Omega_c$, so that the curl of the shift vector has two sources,
\begin{equation}
(\nabla\times \Rv)_z=\underbrace{-\,\Omega_{cv}(\bk)}_{\text{smooth}}
+\underbrace{2\pi\sum_i n_i\,\delta^2(\bk-\bk_i)}_{\text{pointlike, integer}},
\label{eq:curl}
\end{equation}
where $\Omega_{cv}\equiv\Omega_c-\Omega_v$ is the interband Berry curvature, with $\Omega_n=(\nabla\times\mathbf{A}_{nn})_z$ the Berry curvature of band $n$ and $C_n=\frac{1}{2\pi}\int_{\BZ}\Omega_n\,d^2k$ its Chern number. 

When the bands carry Chern numbers, no smooth global gauge exists, and the curl is evaluated patch by patch, in local gauges; since $\Rv$ is gauge invariant and smooth away from the dipole zeros, Eq.~\eqref{eq:curl} holds on the whole torus. Mera and Ozawa wrote the shift vector as a singular connection minus the Berry connection and obtained the Chern number of a transition by counting the zeros of the transition dipole~\cite{MeraOzawa_2022}. Wang and Chang quantised the Berry flux through a patch of the zone together with the winding along its boundary~\cite{wangchang}. Zhu and Alexandradinata proved the Chern-vorticity theorem on the whole torus~\cite{Zhu_2024}. Equation~\eqref{eq:curl} is the differential form of these results. Integrated over a patch of the zone, it gives the first two; integrated over the whole torus, where the curl of the periodic and single-valued field $\Rv$ must vanish, it gives the third,
\begin{equation}
\sum_i n_i=\frac{1}{2\pi}\int_{\BZ}\Omega_{cv}\,d^2k=C_c-C_v\equiv\chi_{cv}.
\label{eq:cv}
\end{equation}
Throughout we distinguish three objects that are all topological in some sense but play different roles: the interband Chern number $\chi_{cv}$, a global invariant; the vortex charges $n_i$, local integers whose sum is fixed by $\chi_{cv}$ but whose number and positions are not; and the vortex sector of the response, defined below, which is linear in the $n_i$. We reserve ``invariant'' for $\chi_{cv}$ and ``charge'' or ``winding'' for $n_i$. For the pair of bands studied below we write $\chi_{cv}=\chi_{32}$.

A vortex is characterised not only by its winding but also by where it sits in the spectrum. If its core $\bk_i$ lies in the resonant region of the pair, so that the transition energy $\omega_{cv}(\bk_i)$ coincides with the absorption onset to within the linewidth, its contribution to the shift current turns on at the onset and it feeds the band-edge peak; we call such a zero a band-edge charge. A zero whose core lies at a transition energy far above the onset carries the same kind of integer but is optically silent in the window of interest. When the core sits at a corner $K$ or $K'$ of the Brillouin zone, or displaced from it by a small fraction of the zone, we call it a corner charge.

Four restrictions delimit what follows. We compute the shift current of a single pair of bands, the resonant pair $2\leftrightarrow3$, which is the full response below the next interband onset ($0.39\eV$ in the phases of Table~\ref{tab:params}, $0.63\eV$ in the biased bilayer of Sec.~\ref{sec:trs}); all spectra of the main text are shown in that window. In Appendix~\ref{app:sectors} we extend the dipole weight of the $2\leftrightarrow3$ pair slightly beyond it, to show the onset of the inactive valley.

We treat one component of the tensor, $\sig$, in the frame defined in Sec.~\ref{sec:model}; the zeros of $r^x_{cv}$ are not those of $r^y_{cv}$, so other components have their own vortices. And we compute the shift current alone: in the Haldane phases, which break time reversal, linearly polarised light also drives a magnetic injection current~\cite{Watanabe_2021} that grows with the relaxation time and is not considered here, whereas in the biased bilayer AB time reversal forbids it and the shift current is the only photocurrent linear in the intensity.

Interactions and excitonic effects are left out. The decomposition itself is general in dimension and symmetry; the point sources are specific to two dimensions, where the zeros of a complex dipole are generically isolated, and a mirror symmetry can force line zeros instead~\cite{Alexandradinata_2024}; in three dimensions the zeros are generically lines, the point charges become vortex lines, and the harmonic part on $T^3$ has three components.

\section{Hodge decomposition of the response}

Equation~\eqref{eq:curl} separates the curl of the shift vector into a quantised part and a smooth one. To carry that separation into $\sig(\omega)$ we must recover the field from its sources, which is what the Helmholtz--Hodge theorem does: on a closed manifold a vector field splits uniquely into a divergence-free (circulating) part, a curl-free (gradient) part and a harmonic part, which on the torus is a constant,
\begin{equation}
\Rv=\underbrace{\hat{\mathbf z}\times\nabla\varphi}_{\text{circulating}}
+\underbrace{\nabla\psi}_{\text{gradient}}
+\underbrace{\Rb}_{\text{harmonic}},
\label{eq:hodge}
\end{equation}
with $\Rb=A_{\BZ}^{-1}\int_{\BZ}\Rv\,d^2k$ the zone average and $A_{\BZ}$ the area of the zone.
Only the circulating part carries curl, so Eqs.~\eqref{eq:hodge} and~\eqref{eq:curl} give a Poisson equation for the stream function, $\nabla^2\varphi=-\Omega_{cv}+2\pi\sum_i n_i\delta^2(\bk-\bk_i)$, and another for the gradient part, $\nabla^2\psi=\varrho(\bk)$ with $\varrho\equiv\nabla\cdot\Rv$ the (gauge-invariant) divergence of the shift vector. The circulating part is split once more according to its two sources, so that the shift vector has four pieces,
\begin{align}
\Rv_{\mathrm{vort}}(\bk)&=\sum_i n_i\,\Kk(\bk-\bk_i),\label{eq:Rvort}\\
\Rv_{\mathrm{Berry}}(\bk)&=-\frac{1}{2\pi}\!\int_{\BZ}\!\Kk(\bk-\bk')\,
\Omega_{cv}(\bk')\,d^2k',\label{eq:Rberry}\\
\Rv_{\mathrm{grad}}(\bk)&=\frac{1}{2\pi}\!\int_{\BZ}\!\Ll(\bk-\bk')\,
\varrho(\bk')\,d^2k',\label{eq:Rgrad}
\end{align}
plus the constant $\Rb$, where
\begin{equation}
\Kk(\bq)=\frac{\hat{\mathbf z}\times\bq}{|\bq|^{2}},
\qquad
\Ll(\bq)=\frac{\bq}{|\bq|^{2}} .
\label{eq:kernels}
\end{equation}
Here $\Kk$ is the vortex kernel, the field of a unit point vortex (the two-dimensional Biot--Savart kernel), and $\Ll$ is the Coulomb kernel, the field of a unit point charge in two dimensions; both derive from the Green function of the two-dimensional Laplacian, $G(\bq)=(2\pi)^{-1}\ln|\bq|$, as $\Ll=2\pi\nabla G$ and $\Kk=\hat{\mathbf z}\times\Ll$, so that $\nabla\cdot\Ll=\nabla\times\Kk=2\pi\,\delta^{2}(\bq)$. 

The kernels are written in their planar form; on the torus every convolution in Eqs.~\eqref{eq:Rvort}--\eqref{eq:Rgrad} is understood with the periodic Green function of the Laplacian. The sum rule~\eqref{eq:cv} makes the total source neutral, which is what makes the Poisson problem solvable on the torus. The two sources separately are not: $-\Omega_{cv}$ and the point charges carry net charge $\mp2\pi\chi_{cv}$, so a uniform neutralising background must be added to each of Eqs.~\eqref{eq:Rvort} and~\eqref{eq:Rberry}. 

The two backgrounds cancel in the sum, so only the split between $\Rv_{\mathrm{vort}}$ and $\Rv_{\mathrm{Berry}}$, and not the total, depends on this choice. Any neutralising distribution would do; the uniform background is the natural choice, the only one that adds no structure, and with it the vortex piece is a sum of translated copies of a single kernel, $\sum_i n_i\Kk_T(\bk-\bk_i)$, with $\Kk_T$ the periodic form of $\Kk$. 

\begin{figure}[t]
\centering
\includegraphics[width=\columnwidth]{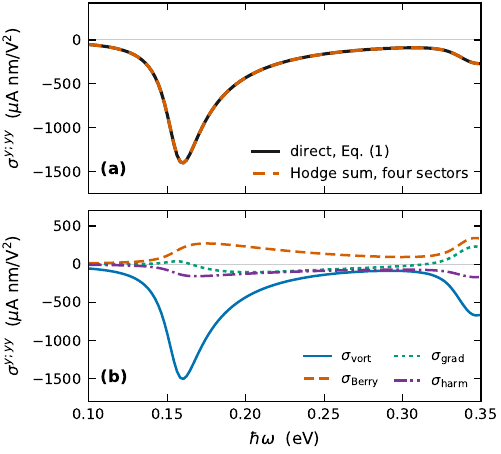}
\caption{Validation of the decomposition, shown for the $\chi_{32}=0$ phase, $2\leftrightarrow3$ pair. (a)~$\sig(\omega)$ obtained by two independent routes: the direct length-gauge integrand of Eq.~\eqref{eq:sigma} (solid) and the Hodge sum of Eq.~\eqref{eq:decomp} (dashed), which overlay; in the Hodge sum $\bar R^y$ is fixed by the additivity of the decomposition and checked against its direct zone average. (b)~Each of the four sectors of Eq.~\eqref{eq:decomp}. Grid $N=2000$, broadening $\eta=10$~meV.}
\label{fig:valid}
\end{figure}

A Hodge decomposition on the Brillouin torus has recently been applied to the Berry curvature itself in the context of the anomalous Hall effect~\cite{WangBraunstein_2026}; here the decomposed field is the shift vector, whose sources include the quantised vortices. Inserting the four pieces into Eq.~\eqref{eq:sigma}, the shift conductivity splits as
\begin{equation}
\sig(\omega)=\svort(\omega)+\sberry(\omega)+\sgrad(\omega)+\sharm(\omega),
\label{eq:decomp}
\end{equation}
which is our central result. The decomposition is an identity: it involves no expansion and holds at every frequency for any nondegenerate band pair of any two-dimensional lattice model. Numerically, with $\bar R^y$ fixed by the additivity of the four sectors and checked against its direct zone average, the identity is verified to $10^{-3}$ of the peak in all five phases of Sec.~\ref{sec:model} (Fig.~\ref{fig:valid} and Table~\ref{tab:cases}).

The harmonic sector is known. Since $\Rb$ is constant,
\begin{align}
\sharm(\omega)&=-\frac{2\pi e^3}{\hbar^2}\,g_s\,\bar R^{y}\,D(\omega),\label{eq:harm}\\
D(\omega)&=\int_{\BZ}\frac{d^2k}{(2\pi)^2}\,|r^y_{cv}|^2\,\delta(\omega_{cv}-\omega),\label{eq:D}
\end{align}
with $D(\omega)$ the dipole weight: it is the zone-averaged shift-vector term of Fregoso, Morimoto and Moore~\cite{Fregoso2017}, resolved in frequency. 

The zone average itself has two parts. Writing $\bk=u_1\mathbf b_1+u_2\mathbf b_2$, with $\mathbf a_i\cdot\mathbf b_j=2\pi\delta_{ij}$, one has $\nabla_{\bk}=\sum_j(\mathbf a_j/2\pi)\,\partial_{u_j}$. Let $w_j$ be the winding of the phase of $r^y_{cv}$ along a closed cycle parallel to $\mathbf b_j$, an integer that jumps by $n_i$ each time the cycle sweeps past a vortex, and $\langle w_j\rangle$ its average over the transverse coordinate; the phase term of the shift vector (Eq.~\eqref{eq:shiftvec}) then averages to $\sum_j\mathbf a_j\langle w_j\rangle$, and
\begin{equation}
\Rb=\langle\mathbf A_{vv}-\mathbf A_{cc}\rangle_{\BZ}+\sum_{j=1,2}\mathbf a_j\,\langle w_j\rangle ,
\label{eq:rbar}
\end{equation}
where the first term is the difference of the hybrid Wannier centres of the two bands and the second depends continuously on the positions of the vortices: moving a charge $n_i$ by a fraction $\Delta s$ of the reciprocal vector $\mathbf b_1$ changes $\langle w_2\rangle$ by $n_i\,\Delta s$ and hence shifts $\Rb$ by $n_i\,\Delta s\,\mathbf a_2$, up to an orientation sign. Only the sum of the two terms is gauge invariant; when $C_v$ or $C_c$ is nonzero, the hybrid Wannier centres are not single-valued, and the split between the two terms depends on the gauge.

For the out-of-plane channel of a slab, the second term is absent, and the zone average reduces to a polarisation difference~\cite{Fregoso2017}; for an in-plane component, it does not, and the harmonic sector is the one place where the vortex configuration enters the response through a single number rather than through the spectrum.

\section{Application: a graphene--Haldane bilayer}
\label{sec:model}

We apply the construction to an AB-stacked graphene bilayer whose lower layer carries a Haldane term~\cite{Haldane1988, McCann_2006, McCann_2013}. In the basis $(A_1,B_1,A_2,B_2)$ the Hamiltonian is
\begin{equation}
H(\mathbf{k}) = \begin{pmatrix}
D_z & t_g f & 0 & \gamma_3 f^* \\
t_g f^* & -D_z & \gamma & 0 \\
0 & \gamma & 0 & t_g f \\
\gamma_3 f & 0 & t_g f^* & 0
\end{pmatrix},
\label{eq:H}
\end{equation}
with $f(\bk)=\sum_j e^{i\bk\cdot\boldsymbol\delta_j}$ the structure factor over the three $A_1\!\to\!B_1$ bonds, $t_g=-2.7\eV$ the intralayer hopping, $\gamma=-0.54\eV$ the vertical dimer coupling between $B_1$ and $A_2$, and $\gamma_3$ the skew interlayer hopping between $A_1$ and $B_2$, which is the trigonal warping. The Haldane diagonal $D_z(\bk)=M-2t_2\sum_i\sin(\bk\cdot\boldsymbol\nu_i)$ (flux phase $\pi/2$) breaks time reversal on the lower layer and, together with the sublattice mass $M$, sets the Chern numbers of the four bands, which we compute on the lattice following Ref.~\cite{Fukui_2005} in the convention of Eq.~\eqref{eq:cv}. 

The Bloch phases of Eq.~\eqref{eq:H} are those of the atomic positions, and the velocity matrix elements are obtained from $\partial_{\bk}H$ with the on-site matrix elements of the position operator neglected; the zeros of $r^y_{cv}$ and their windings depend on this convention for the position operator. We study the resonant middle pair, valence band 2 and conduction band 3, whose transition is the full response below the next interband onset at $0.39\eV$. 

Restricting the resonance to this pair does not reduce the model to two bands: the dipole $r^y_{cv}$, the Berry connections of Eq.~\eqref{eq:shiftvec} and the curvature $\Omega_{cv}$ are those of the eigenstates of the full four-band Hamiltonian, so that the remote bands 1 and 4 enter the shift vector as virtual intermediate states, exactly as in its sum-over-states form~\cite{AversaSipe_1995,sipe2000}; only the absorptive factor $\delta(\omega_{cv}-\omega)$ is restricted to $2\leftrightarrow3$, and the pairs $1\to3$, $2\to4$ and $1\to4$ contribute only above their own onsets.

The frame is fixed as follows. The primitive vectors are $\mathbf a_{1,2}=a(\pm\tfrac12,\tfrac{\sqrt3}{2})$ with $a=2.46$~\AA, so that the nearest-neighbour bond from $A_1$ to $B_1$ points along $+\hat{\mathbf y}$ (the armchair direction); the reciprocal vectors are $\mathbf b_{1,2}=\tfrac{2\pi}{a}(\pm1,\tfrac{1}{\sqrt3})$, the corner we call $K$ is $\mathbf K=\tfrac{2\pi}{a}(\tfrac13,\tfrac{1}{\sqrt3})$ and $K'=\tfrac{2\pi}{a}(\tfrac23,0)$. The light is linearly polarised along $\hat{\mathbf y}$ and the current is measured along the same axis. Since the crystal has $C_3$ but not $C_6$ symmetry, rotating the frame by $60^\circ$, which exchanges the roles of $K$ and $K'$, reverses the sign of $\sig$; all signs below refer to this frame.

\begin{figure}[t]
\centering
\includegraphics[width=\columnwidth]{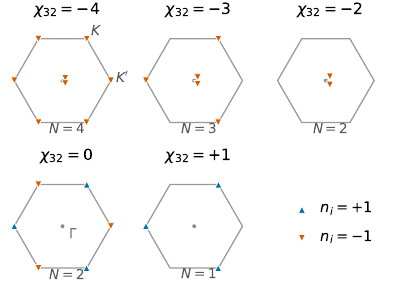}
\caption{Optical vortices of the $y$-polarised $3\!\leftrightarrow\!2$ transition across the five phases, folded into the first Brillouin zone. Up-triangles carry $n_i=+1$ and down-triangles $n_i=-1$; corner charges are drawn at all three equivalent vertices, whereas $N_{\mathrm{vort}}$ counts them per unit cell. The charges sum to $\chi_{32}$ in every panel. $\chi_{32}=-2$ is the only phase with no charge at a zone corner, and the corner charge reverses its winding between $\chi_{32}=-3$ and $\chi_{32}=0$: these two facts, and not the value of $\chi_{32}$, are what the response follows.}
\label{fig:vortices}
\end{figure}

The model is chosen because its interband Chern number $\chi_{32}$ can be swept through five integer values, each with its own vortex configuration (Fig.~\ref{fig:vortices}), while the resonant gap is held fixed: for every phase we choose $(M,t_2)$ so that the band-edge gap of the valley with the smaller gap, which we call the active valley, is $E_g\simeq0.16\eV$ (Table~\ref{tab:params}), which discards the resonant gap as a variable across the phases; we refer to this common value as the matched gap.

Table~\ref{tab:params} lists the band-edge minima. Each valley is centred on a corner of the zone, $K$ or $K'$, and the gap at that corner is not always the minimum of the transition energy in the valley. The trigonal warping adds to the dispersion of the middle pair a term linear in $\bk$ with threefold symmetry, which competes with the curvature of the bands. That curvature shrinks as the gap grows, so in the inactive valley of $\chi_{32}=-4$ and $0$ the warping wins over a wider region and moves the band-edge minimum a short distance away from the corner, along the line from $\Gamma$ to that corner, and a few tens of meV below the corner value; in the other phases the minimum stays at the corner. In the active valley the bands are steeper and the displacement is negligible on the scale of the linewidth, so the gap at the active corner is the resonant gap.

\begin{table}[t]
\centering
\caption{Parameters of the five phases. For each value of the invariant $\chi_{32}$ the pair $(M,t_2)$ is chosen so that the smaller of the two band-edge gaps is $E_g^{\mathrm{min}}\simeq0.16\eV$; the last column names the valley that carries it. $E_g^{K}$ and $E_g^{K'}$ are the minima of the $3\!\leftrightarrow\!2$ transition energy in each valley (see text for their displacement from the corners). The skew interlayer hopping is $\gamma_3=-0.38\eV$ throughout. Chern numbers in the convention of Eq.~\eqref{eq:cv}.}
\label{tab:params}
\begin{ruledtabular}
\begin{tabular}{cccccc}
$\chi_{32}$ & $M$ (eV) & $t_2$ (eV) & $E_g^{K}$ (eV) & $E_g^{K'}$ (eV) & active valley \\
\hline
$-4$ & $0.10$ & $-0.050$ & $0.159$ & $0.333$ & $K$ \\
$-3$ & $0.15$ & $-0.060$ & $0.161$ & $0.356$ & $K$ \\
$-2$ & $0.59$ & $-0.200$ & $0.360$ & $0.163$ & $K'$ \\
$0$ & $0.26$ & $-0.020$ & $0.156$ & $0.336$ & $K$ \\
$+1$ & $0.31$ & $-0.030$ & $0.154$ & $0.355$ & $K$ \\
\end{tabular}
\end{ruledtabular}
\end{table}

\begin{figure}[t]
\centering
\includegraphics[width=\columnwidth]{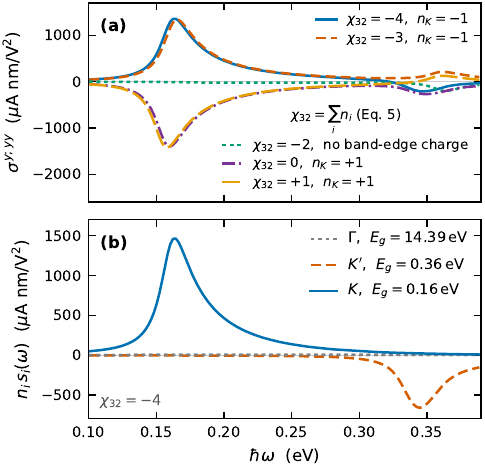}
\caption{(a)~Total $\sig(\omega)$ of the $2\leftrightarrow3$ pair for the five phases at the matched gap $E_g\simeq0.16\eV$, labelled by the invariant $\chi_{32}$ and by the winding $n_K$ of the band-edge charge; phases with a positive peak are listed above the axis, those with a negative peak below. (b)~Per-vortex breakdown of $\svort$ for $\chi_{32}=-4$, Eq.~\eqref{eq:sectors}: the contribution $n_i s_i(\omega)$ of each of the four charges (the two near $\Gamma$ coincide), labelled by position and by the gap at that momentum.}
\label{fig:spectra}
\end{figure}

Figure~\ref{fig:valid} compares the shift current computed from the direct integrand of Eq.~\eqref{eq:sigma} with the sum of the four sectors of Eq.~\eqref{eq:decomp}; the two curves overlay, and Table~\ref{tab:cases} shows that the additivity error is $10^{-3}$ or below in all five phases, with $\bar R^y$ fixed by the same condition and checked against its direct zone average (see Appendix~\ref{app:sectors}). The decomposition therefore reproduces the response, and the vortex sector can be read on its own. The total $\sig(\omega)$ of the five phases, from the direct integrand, is shown in Fig.~\ref{fig:spectra}(a): each phase has a single band-edge peak at the matched gap, of nearly the same magnitude, positive for $\chi_{32}=-4,-3$ and negative for $0,+1$, except $\chi_{32}=-2$, whose response at the matched gap is two orders of magnitude weaker and whose largest feature, at the onset of the $K$ valley, is still several times weaker.

To resolve the vortex sector charge by charge we write
\begin{equation}
    \svort(\omega) = \sum_{i} n_i\, s_i(\omega),\qquad s_i(\omega)\equiv s(\bk_i,\omega),
    \label{eq:sectors}
\end{equation}
with
\begin{equation}
    s(\bk_0,\omega) \equiv -\frac{g_s e^3}{2\pi\hbar^2} \int_{\BZ} \bigl[\Kk_T(\bk-\bk_0)\bigr]_y\, |r_{cv}^y|^2\, \delta(\omega_{cv}-\omega)\, d^2k
    \label{eq:factor_s}
\end{equation}
the spectral profile of a unit charge placed at $\bk_0$, which follows from inserting Eq.~\eqref{eq:Rvort} into Eq.~\eqref{eq:sigma}. Here $\Kk_T$ is the periodic (torus) kernel of Eq.~\eqref{eq:Rvort}, the vortex kernel on the torus built with the periodic Green function (Appendix~\ref{app:green}), so that $\svort=\sum_i n_i s_i$ holds exactly. 

Figure~\ref{fig:spectra}(b) shows these contributions for the phase $\chi_{32}=-4$, which has four charges: one at each corner and two close to $\Gamma$. The charge at the active corner $K$, where the gap is the resonant one, accounts for the entire band-edge peak; the two charges near $\Gamma$, where the transition energy is two orders of magnitude larger, are indistinguishable from zero in the optical window, not because they are distant, since the kernel decays only as $1/q$, but by symmetry: the periodic field of a charge placed at $\Gamma$ vanishes at the zone corners by threefold symmetry, and a charge near $\Gamma$ leaves there only a remnant proportional to its distance from $\Gamma$; and the charge at $K'$ resonates at its own onset, the band edge of the $K'$ valley, visible at the right end of the window. 

Nonetheless, vortices alone are not enough: they must lie at the band edge. Let us analyze the phase $\chi_{32}=-2$ in detail. It is the only phase without a charge at a corner (Fig.~\ref{fig:vortices}); its two charges lie near $\Gamma$, its vortex sector is negligible, and its photocurrent at the matched gap is two orders of magnitude weaker than in the other phases despite its nonzero invariant, and even its largest feature, at the onset of the $K$ valley, is several times weaker (Fig.~\ref{fig:spectra}(a)). The profile of a test charge placed at its band edge is also small, more than an order of magnitude below that of the active valley of the other four phases (Table~\ref{tab:cases}). Thus, the reduction has two causes: the missing charge closes the vortex channel, and the valley geometry of this phase would make any charge there weak.

The origin lies in the character of the band-edge pair. In the four phases with a corner charge, the resonant pair at the active corner is $(A_1,B_2)$: the two sublattices not connected by the nearest-neighbour hopping, intralayer or interlayer. They are directly connected only by $\gamma_3$. At $\gamma_3=0$ the dipole between them vanishes exactly at the corner, so the zero of $r^y_{cv}$, and with it the vortex, is pinned there. The skew hopping $\gamma_3$ lifts the dipole at the corner but cannot remove the zero, whose winding is protected: it displaces the vortex slightly along $\Gamma$--$K$ and switches on its profile. Which pair sits in the middle of the spectrum is decided analytically. At a corner $K_\tau$ ($\tau=+1$ for $K$, $-1$ for $K'$) the structure factor vanishes, so the intralayer hopping and the skew term $\gamma_3 f^*$ drop out of Eq.~\eqref{eq:H}: $A_1$, at energy $D_z(K_\tau)$, and $B_2$, at energy $0$, are decoupled eigenstates, while $B_1$, at energy $-D_z$, and $A_2$, at energy $0$, form the dimer levels $E_\pm=-D_z/2\pm\sqrt{D_z^2/4+\gamma^2}$. Hence $A_1$ and $B_2$ are the middle pair, and the corner hosts a charge if and only if
\begin{equation}
|D_z(K_\tau)|<|\gamma|/\sqrt2 ,
\label{eq:corner}
\end{equation}
which is $\simeq0.38\eV$ here. This criterion reproduces the census of Fig.~\ref{fig:vortices} at all ten corners of the five phases. At the active corner of the four phases with a corner charge, the Haldane mass equals the matched gap and lies well below the threshold. At their inactive corners, it stays just below the threshold in $\chi_{32}=-4$ and $0$, which therefore carry a second charge, and exceeds it in $\chi_{32}=-3$ and $+1$, which do not. In $\chi_{32}=-2$ the mass exceeds the threshold at both corners, by a wide margin at the active one: the upper dimer level, mostly $A_2$, drops into the middle of the spectrum and the resonant pair becomes the intralayer $(B_2,A_2)$ transition, whose dipole does not vanish at the corner and is affected by $\gamma_3$ only through a small $B_1$ component of the band-edge states.

Beyond the charge's position, the peak is governed by its unit profile $s_i(\omega)$. Under an emergent continuous rotational symmetry ($C_\infty$) about the valley, which the unwarped low-energy bands possess, the squared dipole $|r^y_{cv}|^2$ of linearly polarised light contains only the angular harmonics $\ell=0$ and $\ell=\pm2$ about the charge, with no $\ell=\pm1$ component, and the kernel $\Kk$, which carries angular momentum $\ell=\pm1$, projects it to zero: the profile vanishes by an angular-momentum selection rule, and the charge is optically silent regardless of its winding. To activate this channel, the lattice must break the rotational symmetry. In the bilayer, the leading such term is the trigonal warping $\gamma_3$, whose gauge-invariant angular weight is $\ell=\pm3$, the orbital winding of the structure factor combined with the sublattice charges~\cite{noteAB}; this selection rule and its release by warping were derived in Ref.~\cite{noteAB}, and here they appear as properties of the unit profile of a vortex. At first order in $\gamma_3$ it displaces the harmonics of $|r^y_{cv}|^2$ by $\pm3$ and endows the squared dipole with the $\ell=\pm1$ component that $\Kk$ requires, so that $s_i(\omega)$ grows linearly in $\gamma_3$ while the resonant gap stays fixed, since $f(\mathbf K)=0$ (Appendix~\ref{app:warping}).

The profile does not vanish exactly at $\gamma_3=0$. The structure factor is itself trigonally distorted at second order in $\bk$, with the same $\ell=\pm3$ weight as the warping~\cite{noteAB}, so a small residual survives without $\gamma_3$ (Appendix~\ref{app:warping}). The selection rule is exact only in the linearised, unwarped limit. The warping also fixes where the charge sits: its distance to the corner is linear in $\gamma_3$, and vanishes with the warping. Thus the winding provides the integer multiplier $n_i$, and the local geometry, the gap and the trigonal anisotropy, provide the magnitude.

At fixed warping, the sign of the band-edge peak is carried by the winding of the band-edge charge; we keep $\gamma_3$ constant in all five phases. As Fig.~\ref{fig:spectra}(a) and Table~\ref{tab:cases} show, with the resonant gap fixed, the peak magnitude is essentially the same across the four phases with a corner charge, spanning five units of $\chi_{32}$, while its sign follows $n_K$, the winding of the charge at the active corner: the unit profile $s_K(\omega)$ has the same sign in all five phases and reverses sign with $\gamma_3$ (Appendix~\ref{app:warping}), so the sign of the peak is fixed by $n_K$ and reverses with it. The winding follows the local mass. At the corner the resonant gap equals $|D_z(K)|$, and $n_K=\mathrm{sgn}\,D_z(K)$: $D_z(K)=-0.160$ and $-0.162\eV$ for $\chi_{32}=-4,-3$ ($n_K=-1$, positive peak) and $+0.156$ and $+0.154\eV$ for $\chi_{32}=0,+1$ ($n_K=+1$, negative peak). The sign of $D_z(K)$ decides which of $A_1$ and $B_2$ is the valence and which is the conduction state at the corner; exchanging them conjugates the transition dipole and reverses the winding. In the bilayers studied here, this is the mechanism of the sign reversal across the topological transition: the photocurrent reverses because the band-edge charge reverses, and the invariant changes with it only because the same gap closing does both. Sign reversals across topological transitions have been reported in other models~\cite{Tan_2016, Sivianes_2023, LinHsu_2025}, including the Haldane monolayer, for which our criterion predicts no corner charge; whether those reversals follow the winding of a band-edge charge located elsewhere in the zone remains to be examined. 

Figure~\ref{fig:spectra}(a) displays this hierarchy directly. Each spectrum carries two integers: the invariant $\chi_{32}$, which by Eq.~\eqref{eq:cv} is the sum of the windings of all the vortices of the pair, and the winding $n_K$ of the single charge at the band edge. Read by $\chi_{32}$, the figure has no order: the invariant takes five different values, while the spectra fall into two groups of the same shape and opposite sign, plus one weak curve. Read by $n_K$, it has all of it: the two phases with $n_K=-1$ ($\chi_{32}=-4,-3$) give the positive peak, the two with $n_K=+1$ ($\chi_{32}=0,+1$) the negative one, and the only phase without a band-edge charge ($\chi_{32}=-2$) is the weak one, although its invariant lies between the others. The sum rule holds in every phase, the remaining windings sitting at $K'$ and near $\Gamma$, where they are optically silent in this window: what the photocurrent reads is not the sum but the single term at the band edge.

The rule extends to every charge: the sign of its term is $n_i$ times the sign of the unit profile of its valley, and in every system with a corner charge the profile has opposite signs at $K$ and $K'$ (Table~\ref{tab:cases}). In the biased bilayer of Sec.~\ref{sec:trs} this is exact by time reversal, which maps $\bk\to-\bk$, leaves the squared dipole unchanged and reverses the kernel; in the Haldane phases it is a numerical observation rather than a symmetry requirement (in $\chi_{32}=-2$, which has no corner charge, the two profiles have the same sign). Figure~\ref{fig:spectra}(b) shows the two corner charges of the $\chi_{32}=-4$ phase, each resonating at the gap of its own valley: they have the same winding and opposite profiles, so their peaks have opposite signs. The windings themselves follow the mass with a valley factor, $n=\tau\,\mathrm{sgn}\,D_z$, at every corner where Eq.~\eqref{eq:corner} holds, so the two corner terms have the same sign when the mass has the same sign in both valleys and opposite signs when it does not. In $\chi_{32}=-4$, where $D_z(K)$ and $D_z(K')$ have opposite signs, the two valleys compete; in $\chi_{32}=0$, where both are positive, they cooperate; and in the biased bilayer of Sec.~\ref{sec:trs}, where the layer bias is a mass of the same sign in both valleys and time reversal forces $n_{K'}=-n_K$, the two terms are equal and add. 

\begin{table*}[t]
\centering
\caption{Response of the five phases, evaluated at the peak of $\sig$ within the optical window; for the four phases with a corner charge the peak lies at the matched gap, for $\chi_{32}=-2$ it lies at the onset of the $K$ valley near $0.36\eV$, while at the matched gap its response is two orders of magnitude smaller than in the other phases. Conductivities in $\uAnm$, Chern numbers and windings in the convention of Eq.~\eqref{eq:cv}. $N_{\mathrm{vort}}$ is the number of vortices of the $3\!\leftrightarrow\!2$ transition per unit cell, whose windings sum to $\chi_{32}$. $\sharm$ uses the zone average $\bar R^{y}$ fixed by additivity, Eq.~\eqref{eq:S_rbar_closure}. $s_K$ and $s_{K'}$ are the peak values of the profile of Eq.~\eqref{eq:factor_s} for a unit test charge placed at the corner, evaluated locally on a disc around the corner (Appendix~\ref{app:sectors}), each taken at the maximum of its own profile, which lies at the onset of its valley; in the inactive valley of $\chi_{32}=-4$ and $0$, where the band minimum is displaced from the corner, $s_{K'}$ is only indicative. $\mathrm{add}_4=\max|\sig-\sum_X\sigma_X|/\max|\sig|$ is the additivity error over the four sectors and the optical window.}
\label{tab:cases}
\begin{ruledtabular}
\begin{tabular}{cccccccccc}
$\chi_{32}$ & $N_{\mathrm{vort}}$ & $\sig$ & $\svort$ & $\sberry$ & $\sgrad$ & $\sharm$ & $s_K$ & $s_{K'}$ & add$_4$ \\
\hline
$-4$ & 4 & $+1359$ & $+1470$ & $-218$ & $+122$ & $-16.0\pm0.5$ & $-1486$ & $+668$ & $1.0\times10^{-3}$ \\
$-3$ & 3 & $+1338$ & $+1453$ & $-218$ & $+122$ & $-19.8\pm0.4$ & $-1464$ & $+263$ & $0.9\times10^{-3}$ \\
$-2$ & 2 & $-201$ & $+10$ & $+22$ & $-211$ & $-21.9\pm0.1$ & $-326$ & $-19$ & $0.05\times10^{-3}$ \\
$0$ & 2 & $-1401$ & $-1505$ & $+220$ & $+28$ & $-144.2\pm0.4$ & $-1523$ & $+659$ & $0.8\times10^{-3}$ \\
$+1$ & 1 & $-1424$ & $-1525$ & $+220$ & $+27$ & $-144.9\pm0.4$ & $-1546$ & $+246$ & $0.8\times10^{-3}$ \\
\end{tabular}
\end{ruledtabular}
\end{table*}

\section{A time-reversal-symmetric case: the AB bilayer with a layer bias}
\label{sec:trs}

\begin{figure}[t]
\centering
\includegraphics[width=\columnwidth]{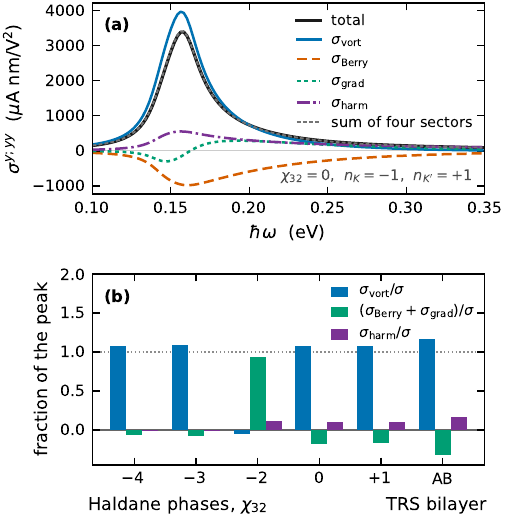}
\caption{(a)~Hodge sectors of $\sig(\omega)$ of the $2\leftrightarrow3$ pair for the time-reversal-symmetric AB bilayer with layer bias $\delta=0.08\eV$ and $\gamma_3=-0.38\eV$ ($\chi_{32}=0$), with the corner gap matched to that of Table~\ref{tab:params}. (b)~Fraction of $\sig$ peak carried by each sector for the five Haldane phases of Table~\ref{tab:cases} and for the biased bilayer.}
\label{fig:trs}
\end{figure}

The decomposition is not tied to Chern phases. For the AB bilayer with a layer bias $\delta$, whose gate-induced shift current has been computed in Ref.~\cite{Zheng_2023} and which is obtained from Eq.~\eqref{eq:H} by replacing the Haldane diagonal with $(-\delta,-\delta,+\delta,+\delta)$, time reversal holds, $\chi_{32}=0$, and the band-edge pair at both corners is again $(A_1,B_2)$. The two corner charges are time-reversal partners with opposite winding and opposite unit profile, $n_{K'}=-n_K$ and $s_{K'}(\omega)=-s_K(\omega)$, so that their vortex contributions add, and $\svort$ alone accounts for the peak, with a small surplus (Fig.~\ref{fig:trs}(a)). The sum of the windings vanishes while the photocurrent does not: the integer that matters is the local one, and at fixed warping the sign is again fixed by $n_K$. 

Reversing the bias, $\delta\to-\delta$, which is equivalent to a spatial inversion and must reverse $\sig$, does so by leaving the two charges in place and exchanging their windings while the unit profiles are unchanged, exactly as the Haldane transition does when $D_z(K)$ changes sign. The harmonic sector is here comparable to that of $\chi_{32}=0,+1$ and much larger than in $\chi_{32}=-4,-3$, the Haldane phases with opposite masses in the two valleys. The first term of Eq.~\eqref{eq:rbar} is the difference of the Wannier centres of the two bands. For an $(A_1,B_2)$ pair these centres sit on sublattices offset in the plane by a bond vector along $\hat{\mathbf y}$, and which of the two is the valence state at each corner is set by the sign of the local mass. The first term therefore approaches that offset when both valleys order the pair alike and nearly cancels when they order it oppositely. Accordingly, $|\bar R^y|$ is close to the bond length in the biased bilayer and in $\chi_{32}=0,+1$, where the mass has the same sign in both valleys, and an order of magnitude smaller in $\chi_{32}=-4,-3$, where it has opposite signs; in $\chi_{32}=-2$ the active pair is intralayer. This is consistent with the harmonic sector being about a tenth of the peak in the first three systems and about a hundredth in the last two (Table~\ref{tab:cases} and Fig.~\ref{fig:trs}(b)).

Figure~\ref{fig:trs}(b) places the six systems side by side: in every one with a corner charge, the vortex sector alone accounts for the peak and the three continuous sectors together remove a small surplus; the harmonic sector follows the valley ordering discussed above. In the phase without a corner charge the peak is carried by the continuous sectors; the harmonic sector is largest in the biased bilayer.

\section{Scope and outlook}

The shift conductivity at a given frequency has no known quantised part, but its shift vector has quantised sources. The Helmholtz-Hodge decomposition on the Brillouin torus turns this into an exact statement about the response: at every frequency, $\sig$ is the sum of a sector linear in the integer vortex charges, a sector driven by the interband Berry curvature, a gradient sector, and the zone-averaged shift vector times the absorption. In every system studied that carries a charge at the band edge, all of them with trigonal warping, the vortex sector alone accounts for the band-edge peak, the three continuous sectors together remove a small surplus, and the zone-averaged term follows the same valley logic as the sign: it is a tenth of the peak or more when the mass has the same sign in both valleys, and about a hundredth when it has opposite signs. The Chern number plays no role in this hierarchy: it fixes the sum of the charges through the sum rule, while the response reads only the charge whose core lies at the band edge. The phase with $\chi_{32}=0$ and the time-reversal-symmetric bilayer behave like the Chern phases, and the only phase that does not is the one whose band-edge pair cannot host a corner charge.

The condition for a band-edge charge is orbital. At $\gamma_3=0$ the two band-edge sublattices of the bilayer, $A_1$ and $B_2$, are not connected by the nearest-neighbour hopping and couple only through the dimer, by an effective hopping $\propto t_g^2f^2/\gamma$ that vanishes quadratically at the corner; the dipole then vanishes linearly there, with unit winding. The skew hopping $\gamma_3$ couples the pair directly, linearly in the distance to the corner, and displaces the zero by an amount linear in $\gamma_3$ without removing it; whether $A_1$ and $B_2$ form the middle pair at a given corner is decided by Eq.~\eqref{eq:corner}. In a gapped monolayer, the two band-edge sublattices are nearest neighbours, their coupling is linear in the distance to the corner and the dipole is finite there, so no corner charge exists and the band-edge shift current lives in the continuous sectors. In a rhombohedral stack of $N$ layers under bias the outer sublattices are coupled as $f^N$ in the chiral limit, so without warping the corner zero has winding $N-1$, and the trigonal warping is expected to split it into $N-1$ unit charges~\cite{CaoWuLouie_2018, Postlewaite_2024}.

Symmetry then decides whether the charge is seen. A charge contributes to $\sig$ only if the squared dipole around it has an $\ell=\pm1$ component, and without warping it has none: it contains only $\ell=0$ and $\ell=\pm2$, so the charge is present but optically silent. The skew hopping $\gamma_3$, which couples the two outer sublattices directly with angular weight $\ell=\pm3$, supplies the missing component: it displaces the charge along $\Gamma$--$K$ and switches on its profile, both linearly in $\gamma_3$. Any perturbation with the same angular weight, $\ell=\pm3$, should act alike.

The sign follows the same logic charge by charge: each corner term is the winding times the unit profile of its valley. The two valleys cooperate when the mass has the same sign in both, as for a layer bias, and compete when it is odd in the valley, as in the phases where the Haldane term dominates the sublattice mass. Time reversal enforces the first case, and reversing the bias exchanges the two windings without moving the charges and reverses the photocurrent, a mechanism that should operate domain by domain in sliding-ferroelectric bilayers. In the models studied here the sign reversal across the topological transition is the reversal of a single band-edge winding, which the invariant merely accompanies; whether the reversals reported in other models~\cite{Tan_2016, Sivianes_2023, LinHsu_2025} admit the same reading remains to be examined. Finally, a large band-edge shift current needs a corner charge, a small gap, since the band-edge response of biased multilayers grows as $1/\delta$~\cite{noteAB}, and an uncancelled trigonal warping; biased rhombohedral multilayers are natural candidates to meet the three, a prediction that remains to be tested by explicit calculation.

\begin{acknowledgments}
E.S.M. acknowledges financial support from Proyecto Interno USM 2026 PI\_LIR\_26\_12.

\emph{Data availability.} The data and codes that support the findings of this article are available from the authors upon reasonable request.

\emph{Use of AI tools.} During the preparation of this work, the authors used Claude (Anthropic; models Fable 5.1 and Opus 5.5, September to October 2026) to write and run plotting scripts on the authors' workstation under their supervision, to check analytical steps against numerical evaluations, to search and verify the literature, and to edit the language of the manuscript. The authors verified every AI-assisted script and result, reviewed and edited all AI-assisted content, and take full responsibility for the content of this publication. No figure was generated or modified by AI image tools; all figures are plots of computed data.
\end{acknowledgments}

\appendix

\section{Decomposition on the Brillouin torus and the kernels}
\label{app:hodge}
\label{app:green}

\subsection{Hodge decomposition}

Fix the light polarisation $b=y$ and the pair of bands $(v,c)$. The shift
vector of Eq.~\eqref{eq:shiftvec} is gauge invariant pointwise, so its curl, its
divergence $\varrho=\nabla\cdot\Rv$ and its zone average are separately gauge
invariant, and so is each sector of the response built from them. By the
Helmholtz--Hodge theorem a periodic field splits uniquely as in
Eq.~\eqref{eq:hodge}, with $\varphi$ and $\psi$ single valued and periodic and
the harmonic part curl and divergence free; on $T^2$ the harmonic fields are the
constants, so the harmonic part is the zone average $\Rb$. The statement
extends to a field with isolated point singularities of the type of
Eq.~\eqref{eq:curl}: the potentials then solve their Poisson equations in the
sense of distributions and, for an isotropic zero, the singular part is carried
entirely by $\varphi$. For an anisotropic zero, $r^y_{cv}\simeq a q_x+ i b q_y$
with $a\neq b$, the phase gradient is not divergence free and $\psi$ also carries
a $1/q$ term, with $\varrho\sim q^{-2}$; the squared dipole
$|r^y_{cv}|^2\propto q^2$ keeps every sector integrable, and in the model studied
the anisotropy of the corner zeros is small. The three pieces are mutually
orthogonal under $\int_{\BZ}\mathbf A\cdot\mathbf B\,d^2k$, understood for the
regularised field with the cores excised, because $\Rv$ itself is not square
integrable, so the splitting is a unique projection and defines \emph{the}
vortex sector of the response.

Taking the curl and divergence of Eq.~\eqref{eq:hodge} gives
\begin{equation}
\nabla^2\varphi=-\,\Omega_{cv}+2\pi\sum_i n_i\,\delta^2(\bk-\bk_i),
\qquad
\nabla^2\psi=\varrho .
\label{eq:S_poisson}
\end{equation}
The Laplacian of a periodic function integrates to zero, so on the torus each
equation is solvable only if its source is neutral,
$\int_{\BZ}\text{(source)}\,d^2k=0$. For the gradient problem neutrality is
automatic, since $\int_{\BZ}\varrho\,d^2k=0$ by the divergence theorem on a
closed surface. For the stream function it is the Chern--vorticity sum rule,
Eq.~\eqref{eq:cv}, which is thus the solvability condition of the problem.

Neither source is neutral by itself: the smooth part integrates to
$-2\pi\chi_{cv}$ and the point charges to $+2\pi\chi_{cv}$. Each half therefore
needs a uniform neutralising background, and the two backgrounds are opposite,
\begin{align}
\nabla^2\varphi_{\mathrm{vort}}&=2\pi\sum_i n_i\,\delta^2(\bk-\bk_i)
-\frac{2\pi\chi_{cv}}{A_{\BZ}},\label{eq:S_phivort}\\
\nabla^2\varphi_{\mathrm{Berry}}&=-\,\Omega_{cv}(\bk)
+\frac{2\pi\chi_{cv}}{A_{\BZ}} .\label{eq:S_phiberry}
\end{align}
They cancel in the sum, so the total field and $\sig(\omega)$ are unchanged;
only the boundary between $\svort$ and $\sberry$ depends on this choice. A
background of any other fixed shape would also give a piece linear in the
charges, but with a kernel that depends on $\bk$ and $\bk_i$ separately.

\subsection{Periodic Green function}

The Green function of the Laplacian on the torus solves
\begin{equation}
\nabla_{\bk}^2\,G_T(\bq)=\delta^2(\bq)-\frac{1}{A_{\BZ}} ,
\label{eq:S_GT}
\end{equation}
the subtracted average of the delta being required for a periodic solution. The Fourier conjugates of $\bk$ are the Bravais
lattice translations $\mathbf T$ of the crystal; with
$G_T(\bq)=\sum_{\mathbf T}\hat G_{\mathbf T}e^{i\mathbf T\cdot\bq}$ and
$\nabla_{\bk}^2\to-|\mathbf T|^2$,
\begin{equation}
\hat G_{\mathbf T}=-\frac{1}{A_{\BZ}|\mathbf T|^{2}}\quad(\mathbf T\neq0),
\qquad \hat G_{\mathbf 0}=0 ,
\label{eq:S_GTfourier}
\end{equation}
the uniform mode being undetermined and not needed. Numerically, we transform
the source, divide by $-|\mathbf T|^2$ away from the origin, set the uniform mode
to zero, apply $\hat{\mathbf z}\times\nabla$ or $\nabla$ in Fourier space and
transform back. Setting it to zero is exact and equivalent to the neutralising
backgrounds of Eqs.~\eqref{eq:S_phivort} and~\eqref{eq:S_phiberry}.

Close to a core, on scales small compared with the zone, $G_T$ reduces to the
planar Green function $G(\bq)=(2\pi)^{-1}\ln|\bq|$, which gives the Coulomb and
vortex kernels of the main text,
\begin{equation}
\Ll(\bq)=2\pi\nabla G=\frac{\bq}{|\bq|^{2}},
\qquad
\Kk(\bq)=\hat{\mathbf z}\times\Ll=\frac{\hat{\mathbf z}\times\bq}{|\bq|^{2}},
\label{eq:S_kernels}
\end{equation}
with $\nabla\cdot\Ll=\nabla\times\Kk=2\pi\delta^2(\bq)$; the numerical work uses
$G_T$ throughout.

The planar kernel cannot be made periodic by a sum over images. The image sum
$\sum_{\mathbf G}\Kk(\bq-\mathbf G)$ over the periods $\mathbf G$ of the torus
has terms falling off as $1/|\mathbf G|$; for a net charge it is only
conditionally convergent, and its value depends on the shape of the summation
region. For a neutral set the leading terms cancel in pairs, but the dipolar
remainder falls off as $1/|\mathbf G|^2$ and is still only conditionally
convergent in two dimensions: it reproduces the torus construction up to a
uniform field that depends on the shape of the summation region, like the
depolarising field of an Ewald sum. An image sum is therefore usable only for a
neutral set.

We checked both statements on an $N=300$ grid. The discrete curl
of the field built with Eq.~\eqref{eq:S_GTfourier} returns
$2\pi n_i\delta^2(\bk-\bk_i)$ at each core plus a uniform background of
$-0.834$~\AA$^{2}$, which differs from $-2\pi/A_{\BZ}$ by less than $10^{-10}$
and integrates to zero net flux. For a neutral pair of charges the torus
construction and the image sum of the planar kernel agree pointwise away from
the cores, with slope $0.9996$ and correlation $0.9994$.

\section{Charges, unit profiles and the zone average}
\label{app:sectors}
\label{app:vortices}
\label{app:warping}

\subsection{Curl of the shift vector and census of the charges}

An optical vortex of the $y$-polarised $c\leftrightarrow v$ transition is a zero
$\bk_i$ of $r^{y}_{cv}$ with nonzero circulation; the circulation of $\Rv$
around an infinitesimal loop $\partial D_i$ enclosing it, traversed
anticlockwise in the $(k_x,k_y)$ plane, is an integer multiple of $2\pi$,
\begin{equation}
n_i=\frac{1}{2\pi}\oint_{\partial D_i}\nabla\arg r^{y}_{cv}\cdot d\bk
=\frac{1}{2\pi}\oint_{\partial D_i}\Rv\cdot d\bk\in\mathbb Z .
\label{eq:S_circ}
\end{equation}
The second equality holds because the connection difference is regular at the
core and its circulation vanishes with the loop.

\emph{Connection part.} With $\mathbf A_{nn}=i\langle u_n|\nabla u_n\rangle$ and
$\Omega_n=(\nabla\times\mathbf A_{nn})_z$,
\begin{equation}
\bigl[\nabla\times(\mathbf A_{vv}-\mathbf A_{cc})\bigr]_z
=\Omega_v-\Omega_c=-\,\Omega_{cv}(\bk),
\label{eq:S_curlA}
\end{equation}
a smooth function of $\bk$ wherever the pair is nondegenerate.

\emph{Phase part.} In a gauge that is smooth near a given point, the curl of the
phase gradient vanishes there unless the dipole vanishes. When $C_v$ or $C_c$ is
nonzero no smooth periodic gauge exists on the whole torus, and the curl is
evaluated patch by patch, in local gauges. The gauge singularities of
$\arg r^{y}_{cv}$ and of $\mathbf A_{vv}-\mathbf A_{cc}$ cancel in $\Rv$, which
is gauge invariant and smooth away from the dipole zeros, so the winding $n_i$
defined by Eq.~\eqref{eq:S_circ} is gauge invariant. At each zero the phase winds by $2\pi n_i$, and by
Stokes' theorem the curl is the distribution
\begin{equation}
\bigl[\nabla\times\nabla\arg r^{y}_{cv}\bigr]_z
=2\pi\sum_i n_i\,\delta^2(\bk-\bk_i).
\label{eq:S_curlphase}
\end{equation}
Equations~\eqref{eq:S_curlA} and~\eqref{eq:S_curlphase} hold separately only
patch by patch. Adding them in each patch gives Eq.~\eqref{eq:curl}, which
involves only $\Rv$ and holds on the whole torus. The convention for
$\mathbf A_{nn}$ fixes the relative sign of the two sources, and with it the sign
of each $n_i$ and of $\chi_{cv}$.

The vortices are located in two stages. On a coarse grid the winding of
$\arg r^{y}_{cv}$ is accumulated around every plaquette from the link phases of
the dipole combined with the overlaps of the band states at neighbouring points,
which makes the result gauge invariant; plaquettes with nonzero winding flag a core. Each core is then
refined by Nelder--Mead minimisation of $|r^{y}_{cv}|$, a gauge-invariant
scalar, to a precision well below the grid spacing, and its charge is measured
by the circulation of Eq.~\eqref{eq:S_circ} on a small loop around the refined
position; a zero with zero winding is not a vortex. Pairs of opposite charge
closer than a couple of grid spacings are flagged, since neither the refinement
nor the unit profile below is meaningful for a pair the grid cannot separate. In
the systems studied the only such pair is the grid splitting of a zero of zero
winding that symmetry places exactly at $\Gamma$, where the velocity vanishes,
at a transition energy two orders of magnitude above the optical window. The
charges near $\Gamma$ in Fig.~\ref{fig:vortices} are distinct: they carry
windings of equal sign, lie $0.13$--$0.20$~\AA$^{-1}$ from $\Gamma$ and survive
the refinement. The sum rule~\eqref{eq:cv} holds in every system, an
independent check of the census, of the Chern numbers, computed on the lattice
following Ref.~\cite{Fukui_2005}, and of the grid resolution.

\subsection{Unit profiles}

Since $\Rv_{\mathrm{vort}}$ is linear in the charges, so is its contribution to
the response, the integer sector of Eq.~\eqref{eq:sectors}, with
\begin{equation}
s(\bk_0,\omega)=-\frac{g_se^3}{2\pi\hbar^2}\int_{\BZ}
\bigl[\Kk_T(\bk-\bk_0)\bigr]_y\,|r^{y}_{cv}|^2\,\delta(\omega_{cv}-\omega)\,d^2k
\label{eq:S_profile}
\end{equation}
the unit profile of a unit charge placed at $\bk_0$, and $\Kk_T$ the periodic kernel of Appendix~\ref{app:green}. The profile depends on
the system and on the point, not on the charge.

With the periodic kernel on the full grid, Eq.~\eqref{eq:S_profile} gives the
per-charge curves of Fig.~\ref{fig:spectra}(b). Evaluated at the corner rather
than at an actual zero, it gives the unit profiles $s_K$ and $s_{K'}$ of a test
charge, quoted in Table~\ref{tab:cases}. These are evaluated with the planar
kernel on a polar grid centred on the corner (a disc of radius
$0.6$~\AA$^{-1}$), which resolves the $1/q$ kernel near the core and makes the
angular integral explicit. This local approximation agrees with the periodic
construction to within $1.5\%$ at the active corner, so that at the peak $n_Ks_K$
reproduces $\svort$ to the same accuracy in the four phases with a band-edge
charge (Table~\ref{tab:cases}).

\subsection{The harmonic sector and the zone average}

\begin{figure}[t]
\centering
\includegraphics[width=0.95\columnwidth]{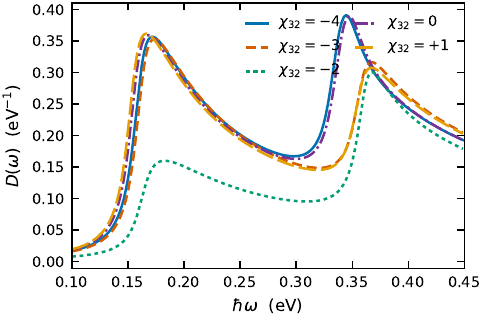}
\caption{Dipole weight $D(\omega)$ of Eq.~\eqref{eq:S_D} for the
$3\!\leftrightarrow\!2$ transition in the five phases ($N=2000$, $\eta=10$~meV). Each phase shows two onsets, one per valley, at its
two band-edge gaps. At the peak of $\sig$ the weight agrees to within one per
cent in the four phases with a band-edge charge, whereas in $\chi_{32}=-2$ it is about a third of that value at the matched gap and about a seventh smaller at its own peak near $0.36\eV$; neither accounts for the two orders of magnitude by which its response at the matched gap falls short, so the drop is not an absorption effect. Above $0.39\eV$, the
onset of the next pair, the $2\leftrightarrow3$ pair is no longer the full
response.}
\label{fig:SM_D}
\end{figure}

The harmonic piece is constant, so it factors out of the response and multiplies
the absorption as in Eq.~\eqref{eq:harm}, with
\begin{equation}
D(\omega)=\int_{\BZ}\frac{d^2k}{(2\pi)^2}\,|r^{y}_{cv}|^2\,
\delta(\omega_{cv}-\omega)
\label{eq:S_D}
\end{equation}
the dipole weight, a dipole-weighted joint density of states
independent of the sign structure of the shift vector (Fig.~\ref{fig:SM_D}).

The zone average has two parts. Integrating Eq.~\eqref{eq:shiftvec} over the
zone, the connection difference gives the difference of the hybrid Wannier
centres of the two bands. For the phase term we write
$\bk=u_1\mathbf b_1+u_2\mathbf b_2$, with $\mathbf a_i\cdot\mathbf b_j=2\pi\delta_{ij}$,
so that $\nabla_{\bk}=\sum_j(\mathbf a_j/2\pi)\,\partial_{u_j}$. Let $w_j$ be the
winding of the phase of $r^{y}_{cv}$ along a closed cycle parallel to
$\mathbf b_j$, an integer that jumps by $n_i$ whenever the cycle sweeps past a
vortex, and $\langle w_j\rangle$ its average over the transverse coordinate.
The zone average of the phase gradient is then $\sum_j\mathbf a_j\langle w_j\rangle$,
which gives Eq.~\eqref{eq:rbar}.

We fix $\bar R^{y}$ by the additivity of the four sectors,
\begin{equation}
\begin{split}
\bar R^{y}=-\frac{\hbar^{2}}{2\pi e^{3}g_s}\Bigl\langle&\frac{1}{D(\omega)}\bigl[\sig(\omega)-\svort(\omega)\\
&-\sberry(\omega)-\sgrad(\omega)\bigr]\Bigr\rangle_{\omega},
\end{split}
\label{eq:S_rbar_closure}
\end{equation}
averaged over the optical window where $D(\omega)$ exceeds a tenth of its
maximum. Since the harmonic sector, Eq.~\eqref{eq:harm}, is the only one
proportional to $D(\omega)$ with a constant coefficient, the bracket is
independent of $\omega$ if the other three sectors are exact. Its residual
variation over the window, below $3\%$ of $\sharm$, is the uncertainty quoted for $\sharm$ in the main text. The direct zone average of
$R^y$ on the grid, defined below Eq.~\eqref{eq:hodge}, does not rely on link
phases and provides an independent check: the two values agree to within two per
cent in four phases and six per cent in $\chi_{32}=-2$, a difference that
changes $\sharm$ by less than one per cent of the peak.

The residual of the four-sector sum with this $\bar R^{y}$ has two components:
a feature at the resonant onset, where the integrand is sharpest and the numerical error of the
three computed sectors is largest, and a small oscillation above $0.2\eV$ of numerical origin,
the Gibbs ringing of the $1/q$ core of the vortex field on the grid, whose period
scales with the grid spacing and whose amplitude is a ten-thousandth of the peak;
the phase without a band-edge charge shows neither.
The additivity error over the whole optical window
is at or below $10^{-3}$ in the five Haldane phases at $N=2000$ and $\eta=10$~meV, and $8\times10^{-3}$ in the biased bilayer, where the residual is concentrated at the onset. With the
direct zone average of $R^y$ instead of the value fixed by additivity, it stays at the
$10^{-3}$ level in four phases and reaches $7\times10^{-3}$ in $\chi_{32}=-2$,
whose peak is the smallest.

\subsection{Unit profile under trigonal warping}

\begin{figure}[t]
\centering
\includegraphics[width=\columnwidth]{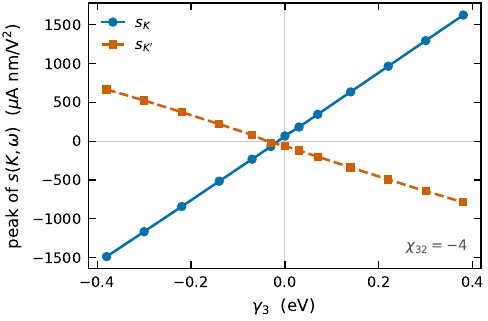}
\caption{Release of the unit profile of the band-edge charge by trigonal
warping, phase $\chi_{32}=-4$ with $(M,t_2)$ as in the main text. Peak of the
unit profile, Eq.~\eqref{eq:S_profile}, of a test charge at $K$ ($s_K$) and at
$K'$ ($s_{K'}$), as a function of $\gamma_3$ from $-0.38$ to $+0.38\eV$. The
profile is linear in $\gamma_3$, with a small offset, and reverses sign with it,
as Eq.~\eqref{eq:S_mu1} requires. The intercept at $\gamma_3=0$ is $4.5\%$ of
the value at $\gamma_3=-0.38\eV$, is local to the valley (unchanged for
integration radii between $0.10$ and $0.60$~\AA$^{-1}$) and does not come from
the Haldane term.}
\label{fig:SM_gamma3}
\end{figure}

Write the kernel in polar coordinates about the core,
$\bq=q(\cos\theta,\sin\theta)$. From Eq.~\eqref{eq:S_kernels},
\begin{equation}
\mathcal K_x=-\frac{\sin\theta}{q},\qquad
\mathcal K_y=\frac{\cos\theta}{q},\qquad
\mathcal K_x+i\mathcal K_y=\frac{i\,e^{i\theta}}{q},
\label{eq:S_kernel_polar}
\end{equation}
so the kernel is an $\ell=\pm1$ object. We expand the squared dipole in angular
harmonics about the same point,
\begin{equation}
\bigl|r^{y}_{cv}\bigr|^2=\sum_{\ell}\mu_\ell(q)\,e^{i\ell\theta},
\qquad \mu_{-\ell}=\mu_\ell^{*},
\label{eq:S_mu}
\end{equation}
where the second relation holds because the squared dipole is real. On a circular
resonance contour centred at the charge, the delta function of
Eq.~\eqref{eq:S_profile} fixes the radius and leaves an angular average,
\begin{equation}
\int_0^{2\pi}\!\!d\theta\;\mathcal K_y\,\bigl|r^{y}_{cv}\bigr|^2
=\frac{2\pi}{q}\,\mathrm{Re}\,\mu_{+1}(q) .
\label{eq:S_angular}
\end{equation}
The profile $s_i(\omega)$ is thus the $\ell=+1$ projection of the squared dipole. Being a signed angular asymmetry, it has no definite sign; its relative
sign at $K$ and $K'$ is discussed in Sec.~\ref{sec:model}. Under the emergent
rotational symmetry of the linearised, unwarped cones the squared dipole contains only
the harmonics $\ell=0$ and $\ell=\pm2$, so $\mu_{\pm1}=0$ and the profile
vanishes identically.

The skew hopping $\gamma_3$ carries angular weight $\ell=\pm3$ and, at first
order, endows the squared dipole with the $\ell=\pm1$ component that the kernel
requires. Writing the dipole to this order as
$r^{y}_{cv}=a_{+1}e^{i\theta}+a_{-1}e^{-i\theta}
+\gamma_3(b_{+2}e^{2i\theta}+b_{-2}e^{-2i\theta})+\dots$, one finds
\begin{equation}
\mu_{+1}\bigl[|r^{y}_{cv}|^2\bigr]
=\gamma_3\bigl(a^{*}_{+1}b_{+2}+a_{-1}b^{*}_{-2}\bigr)+O(\gamma_3^2),
\label{eq:S_mu1}
\end{equation}
linear and odd in $\gamma_3$; the derivation is given in Ref.~\cite{noteAB},
where the selection rule was obtained for the band-edge shift current itself.
With warping the resonance contour also acquires a threefold distortion that
contributes at the same order, so Eq.~\eqref{eq:S_angular} gives the leading
angular-momentum argument, not the full first-order coefficient. The profile of
the band-edge charge therefore grows linearly with the warping and reverses sign
with it (Fig.~\ref{fig:SM_gamma3}). Along the sweep the gaps at $K$ and $K'$ do
not change, since $f(\mathbf K)=0$ removes $\gamma_3$ from the Hamiltonian at
the corners, so the growth of the profile is an angular-momentum release and
not a gap effect. The small intercept at $\gamma_3=0$ is not a Haldane
contribution: it comes from the trigonal distortion of the structure factor at
second order in $\bk$, which carries the same $\ell=\pm3$ weight as $\gamma_3$
and is absent only in the strictly linearised model~\cite{noteAB}.

The same hopping fixes the position of the charge. Sweeping $\gamma_3$ from
$-0.38$ to $+0.38\eV$ in the phase $\chi_{32}=-4$, the zero of $r^{y}_{cv}$
moves only along the radial ($\Gamma$--$K$) direction, by an amount linear in
$\gamma_3$ to numerical accuracy, $0.0061$~\AA$^{-1}$ at $\gamma_3=-0.38\eV$, and
sits exactly at the $C_3$ point when the warping is off; its winding,
$n_K=-1$, is unchanged throughout the sweep. At the physical value of
$\gamma_3$ the displacement is a small fraction of a per cent of
$|\mathbf b_1|$, hence the term corner charge, but the profile of a charge near
a band edge is sensitive to its position at that level, and the census
therefore refines the zeros well below the grid spacing.

\bibliography{refs}

\begin{thebibliography}{29}%
\makeatletter
\providecommand \@ifxundefined [1]{%
 \@ifx{#1\undefined}
}%
\providecommand \@ifnum [1]{%
 \ifnum #1\expandafter \@firstoftwo
 \else \expandafter \@secondoftwo
 \fi
}%
\providecommand \@ifx [1]{%
 \ifx #1\expandafter \@firstoftwo
 \else \expandafter \@secondoftwo
 \fi
}%
\providecommand \natexlab [1]{#1}%
\providecommand \enquote  [1]{``#1''}%
\providecommand \bibnamefont  [1]{#1}%
\providecommand \bibfnamefont [1]{#1}%
\providecommand \citenamefont [1]{#1}%
\providecommand \href@noop [0]{\@secondoftwo}%
\providecommand \href [0]{\begingroup \@sanitize@url \@href}%
\providecommand \@href[1]{\@@startlink{#1}\@@href}%
\providecommand \@@href[1]{\endgroup#1\@@endlink}%
\providecommand \@sanitize@url [0]{\catcode `\\12\catcode `\$12\catcode
  `\&12\catcode `\#12\catcode `\^12\catcode `\_12\catcode `\%12\relax}%
\providecommand \@@startlink[1]{}%
\providecommand \@@endlink[0]{}%
\providecommand \url  [0]{\begingroup\@sanitize@url \@url }%
\providecommand \@url [1]{\endgroup\@href {#1}{\urlprefix }}%
\providecommand \urlprefix  [0]{URL }%
\providecommand \Eprint [0]{\href }%
\providecommand \doibase [0]{https://doi.org/}%
\providecommand \selectlanguage [0]{\@gobble}%
\providecommand \bibinfo  [0]{\@secondoftwo}%
\providecommand \bibfield  [0]{\@secondoftwo}%
\providecommand \translation [1]{[#1]}%
\providecommand \BibitemOpen [0]{}%
\providecommand \bibitemStop [0]{}%
\providecommand \bibitemNoStop [0]{.\EOS\space}%
\providecommand \EOS [0]{\spacefactor3000\relax}%
\providecommand \BibitemShut  [1]{\csname bibitem#1\endcsname}%
\let\auto@bib@innerbib\@empty
\bibitem [{\citenamefont {de~Juan}\ \emph {et~al.}(2017)\citenamefont
  {de~Juan}, \citenamefont {Grushin}, \citenamefont {Morimoto},\ and\
  \citenamefont {Moore}}]{deJuan_2017}%
  \BibitemOpen
  \bibfield  {author} {\bibinfo {author} {\bibfnamefont {F.}~\bibnamefont
  {de~Juan}}, \bibinfo {author} {\bibfnamefont {A.~G.}\ \bibnamefont
  {Grushin}}, \bibinfo {author} {\bibfnamefont {T.}~\bibnamefont {Morimoto}},\
  and\ \bibinfo {author} {\bibfnamefont {J.~E.}\ \bibnamefont {Moore}},\
  }\bibfield  {title} {\bibinfo {title} {Quantized circular photogalvanic
  effect in {Weyl} semimetals},\ }\href {https://doi.org/10.1038/ncomms15995}
  {\bibfield  {journal} {\bibinfo  {journal} {Nat. Commun.}\ }\textbf {\bibinfo
  {volume} {8}},\ \bibinfo {pages} {15995} (\bibinfo {year}
  {2017})}\BibitemShut {NoStop}%
\bibitem [{\citenamefont {Ahn}\ \emph {et~al.}(2020)\citenamefont {Ahn},
  \citenamefont {Guo},\ and\ \citenamefont {Nagaosa}}]{Ahn_2020}%
  \BibitemOpen
  \bibfield  {author} {\bibinfo {author} {\bibfnamefont {J.}~\bibnamefont
  {Ahn}}, \bibinfo {author} {\bibfnamefont {G.-Y.}\ \bibnamefont {Guo}},\ and\
  \bibinfo {author} {\bibfnamefont {N.}~\bibnamefont {Nagaosa}},\ }\bibfield
  {title} {\bibinfo {title} {Low-frequency divergence and quantum geometry of
  the bulk photovoltaic effect in topological semimetals},\ }\href
  {https://doi.org/10.1103/PhysRevX.10.041041} {\bibfield  {journal} {\bibinfo
  {journal} {Phys. Rev. X}\ }\textbf {\bibinfo {volume} {10}},\ \bibinfo
  {pages} {041041} (\bibinfo {year} {2020})}\BibitemShut {NoStop}%
\bibitem [{\citenamefont {Alexandradinata}(2024)}]{Alexandradinata_2024}%
  \BibitemOpen
  \bibfield  {author} {\bibinfo {author} {\bibfnamefont {A.}~\bibnamefont
  {Alexandradinata}},\ }\bibfield  {title} {\bibinfo {title} {Quantization of
  intraband and interband {Berry} phases in the shift current},\ }\href
  {https://doi.org/10.1103/PhysRevB.110.075159} {\bibfield  {journal} {\bibinfo
   {journal} {Phys. Rev. B}\ }\textbf {\bibinfo {volume} {110}},\ \bibinfo
  {pages} {075159} (\bibinfo {year} {2024})},\ \Eprint
  {https://arxiv.org/abs/2203.11225} {arXiv:2203.11225 [cond-mat.mes-hall]}
  \BibitemShut {NoStop}%
\bibitem [{\citenamefont {Wang}\ and\ \citenamefont {Chang}(2026)}]{wangchang}%
  \BibitemOpen
  \bibfield  {author} {\bibinfo {author} {\bibfnamefont {H.}~\bibnamefont
  {Wang}}\ and\ \bibinfo {author} {\bibfnamefont {K.}~\bibnamefont {Chang}},\
  }\bibfield  {title} {\bibinfo {title} {Geodesic nature and quantization of
  shift vector},\ }\href {https://doi.org/10.1088/0256-307X/43/2/020703}
  {\bibfield  {journal} {\bibinfo  {journal} {Chin. Phys. Lett.}\ }\textbf
  {\bibinfo {volume} {43}},\ \bibinfo {pages} {020703} (\bibinfo {year}
  {2026})},\ \Eprint {https://arxiv.org/abs/2405.13355} {arXiv:2405.13355
  [cond-mat.mtrl-sci]} \BibitemShut {NoStop}%
\bibitem [{\citenamefont {Jankowski}\ and\ \citenamefont
  {Slager}(2024)}]{Jankowski_2024}%
  \BibitemOpen
  \bibfield  {author} {\bibinfo {author} {\bibfnamefont {W.~J.}\ \bibnamefont
  {Jankowski}}\ and\ \bibinfo {author} {\bibfnamefont {R.-J.}\ \bibnamefont
  {Slager}},\ }\bibfield  {title} {\bibinfo {title} {Quantized integrated shift
  effect in multigap topological phases},\ }\href
  {https://doi.org/10.1103/PhysRevLett.133.186601} {\bibfield  {journal}
  {\bibinfo  {journal} {Phys. Rev. Lett.}\ }\textbf {\bibinfo {volume} {133}},\
  \bibinfo {pages} {186601} (\bibinfo {year} {2024})},\ \Eprint
  {https://arxiv.org/abs/2402.13245} {arXiv:2402.13245 [cond-mat.mes-hall]}
  \BibitemShut {NoStop}%
\bibitem [{\citenamefont {Tan}\ and\ \citenamefont {Rappe}(2016)}]{Tan_2016}%
  \BibitemOpen
  \bibfield  {author} {\bibinfo {author} {\bibfnamefont {L.~Z.}\ \bibnamefont
  {Tan}}\ and\ \bibinfo {author} {\bibfnamefont {A.~M.}\ \bibnamefont
  {Rappe}},\ }\bibfield  {title} {\bibinfo {title} {Enhancement of the bulk
  photovoltaic effect in topological insulators},\ }\href
  {https://doi.org/10.1103/PhysRevLett.116.237402} {\bibfield  {journal}
  {\bibinfo  {journal} {Phys. Rev. Lett.}\ }\textbf {\bibinfo {volume} {116}},\
  \bibinfo {pages} {237402} (\bibinfo {year} {2016})}\BibitemShut {NoStop}%
\bibitem [{\citenamefont {Sivianes}\ and\ \citenamefont
  {{Iba\~nez-Azpiroz}}(2023)}]{Sivianes_2023}%
  \BibitemOpen
  \bibfield  {author} {\bibinfo {author} {\bibfnamefont {J.}~\bibnamefont
  {Sivianes}}\ and\ \bibinfo {author} {\bibfnamefont {J.}~\bibnamefont
  {{Iba\~nez-Azpiroz}}},\ }\bibfield  {title} {\bibinfo {title} {Shift
  photoconductivity in the {Haldane} model},\ }\href
  {https://doi.org/10.1103/PhysRevB.108.155419} {\bibfield  {journal} {\bibinfo
   {journal} {Phys. Rev. B}\ }\textbf {\bibinfo {volume} {108}},\ \bibinfo
  {pages} {155419} (\bibinfo {year} {2023})},\ \Eprint
  {https://arxiv.org/abs/2305.17035} {arXiv:2305.17035 [cond-mat.mes-hall]}
  \BibitemShut {NoStop}%
\bibitem [{\citenamefont {Lin}\ and\ \citenamefont {Hsu}(2026)}]{LinHsu_2025}%
  \BibitemOpen
  \bibfield  {author} {\bibinfo {author} {\bibfnamefont {B.-X.}\ \bibnamefont
  {Lin}}\ and\ \bibinfo {author} {\bibfnamefont {H.-C.}\ \bibnamefont {Hsu}},\
  }\bibfield  {title} {\bibinfo {title} {Bulk photovoltaic effects in
  mirror-time symmetric systems},\ }\href {https://doi.org/10.1103/xqyz-wtc7}
  {\bibfield  {journal} {\bibinfo  {journal} {Phys. Rev. B}\ }\textbf {\bibinfo
  {volume} {113}},\ \bibinfo {pages} {085424} (\bibinfo {year} {2026})},\
  \Eprint {https://arxiv.org/abs/2508.12414} {arXiv:2508.12414
  [cond-mat.mes-hall]} \BibitemShut {NoStop}%
\bibitem [{\citenamefont {Mera}\ and\ \citenamefont
  {Ozawa}(2022)}]{MeraOzawa_2022}%
  \BibitemOpen
  \bibfield  {author} {\bibinfo {author} {\bibfnamefont {B.}~\bibnamefont
  {Mera}}\ and\ \bibinfo {author} {\bibfnamefont {T.}~\bibnamefont {Ozawa}},\
  }\bibfield  {title} {\bibinfo {title} {Singular-connection approach to
  topological phases and resonant optical responses},\ }\href
  {https://doi.org/10.1103/PhysRevB.106.245134} {\bibfield  {journal} {\bibinfo
   {journal} {Phys. Rev. B}\ }\textbf {\bibinfo {volume} {106}},\ \bibinfo
  {pages} {245134} (\bibinfo {year} {2022})},\ \Eprint
  {https://arxiv.org/abs/2210.06844} {arXiv:2210.06844 [cond-mat.mes-hall]}
  \BibitemShut {NoStop}%
\bibitem [{\citenamefont {Zhu}\ and\ \citenamefont
  {Alexandradinata}(2024)}]{Zhu_2024}%
  \BibitemOpen
  \bibfield  {author} {\bibinfo {author} {\bibfnamefont {P.}~\bibnamefont
  {Zhu}}\ and\ \bibinfo {author} {\bibfnamefont {A.}~\bibnamefont
  {Alexandradinata}},\ }\bibfield  {title} {\bibinfo {title} {Anomalous shift
  and optical vorticity in the steady photovoltaic current},\ }\href
  {https://doi.org/10.1103/PhysRevB.110.115108} {\bibfield  {journal} {\bibinfo
   {journal} {Phys. Rev. B}\ }\textbf {\bibinfo {volume} {110}},\ \bibinfo
  {pages} {115108} (\bibinfo {year} {2024})},\ \Eprint
  {https://arxiv.org/abs/2308.08596} {arXiv:2308.08596 [cond-mat.mes-hall]}
  \BibitemShut {NoStop}%
\bibitem [{\citenamefont {Sukhachov}\ \emph {et~al.}(2026)\citenamefont
  {Sukhachov}, \citenamefont {Zhu}, \citenamefont {Banyas}, \citenamefont
  {Tan},\ and\ \citenamefont {Alexandradinata}}]{Sukhachov_2026}%
  \BibitemOpen
  \bibfield  {author} {\bibinfo {author} {\bibfnamefont {P.}~\bibnamefont
  {Sukhachov}}, \bibinfo {author} {\bibfnamefont {P.}~\bibnamefont {Zhu}},
  \bibinfo {author} {\bibfnamefont {E.}~\bibnamefont {Banyas}}, \bibinfo
  {author} {\bibfnamefont {L.~Z.}\ \bibnamefont {Tan}},\ and\ \bibinfo {author}
  {\bibfnamefont {A.}~\bibnamefont {Alexandradinata}},\ }\href@noop {}
  {\bibinfo {title} {Vortex-enhanced photovoltaic current in disordered
  topological materials}} (\bibinfo {year} {2026}),\ \Eprint
  {https://arxiv.org/abs/2606.28509} {arXiv:2606.28509 [cond-mat.str-el]}
  \BibitemShut {NoStop}%
\bibitem [{\citenamefont {P\'erez~Riffo}\ \emph {et~al.}(2026)\citenamefont
  {P\'erez~Riffo}, \citenamefont {Castro~Schnaidt}, \citenamefont {Chico},
  \citenamefont {Foa~Torres},\ and\ \citenamefont {Su\'arez~Morell}}]{noteAB}%
  \BibitemOpen
  \bibfield  {author} {\bibinfo {author} {\bibfnamefont {F.}~\bibnamefont
  {P\'erez~Riffo}}, \bibinfo {author} {\bibfnamefont {M.}~\bibnamefont
  {Castro~Schnaidt}}, \bibinfo {author} {\bibfnamefont {L.}~\bibnamefont
  {Chico}}, \bibinfo {author} {\bibfnamefont {L.~E.~F.}\ \bibnamefont
  {Foa~Torres}},\ and\ \bibinfo {author} {\bibfnamefont {E.}~\bibnamefont
  {Su\'arez~Morell}},\ }\href@noop {} {\bibinfo {title} {Symmetry selection
  rule for the band-edge shift current in two dimensions}} (\bibinfo {year}
  {2026}),\ \bibinfo {note} {companion paper},\ \Eprint
  {https://arxiv.org/abs/2607.28927} {arXiv:2607.28927 [cond-mat.mes-hall]}
  \BibitemShut {NoStop}%
\bibitem [{\citenamefont {Aversa}\ and\ \citenamefont
  {Sipe}(1995)}]{AversaSipe_1995}%
  \BibitemOpen
  \bibfield  {author} {\bibinfo {author} {\bibfnamefont {C.}~\bibnamefont
  {Aversa}}\ and\ \bibinfo {author} {\bibfnamefont {J.~E.}\ \bibnamefont
  {Sipe}},\ }\bibfield  {title} {\bibinfo {title} {Nonlinear optical
  susceptibilities of semiconductors: {Results} with a length-gauge analysis},\
  }\href {https://doi.org/10.1103/PhysRevB.52.14636} {\bibfield  {journal}
  {\bibinfo  {journal} {Phys. Rev. B}\ }\textbf {\bibinfo {volume} {52}},\
  \bibinfo {pages} {14636} (\bibinfo {year} {1995})}\BibitemShut {NoStop}%
\bibitem [{\citenamefont {Sipe}\ and\ \citenamefont
  {Shkrebtii}(2000)}]{sipe2000}%
  \BibitemOpen
  \bibfield  {author} {\bibinfo {author} {\bibfnamefont {J.~E.}\ \bibnamefont
  {Sipe}}\ and\ \bibinfo {author} {\bibfnamefont {A.~I.}\ \bibnamefont
  {Shkrebtii}},\ }\bibfield  {title} {\bibinfo {title} {Second-order optical
  response in semiconductors},\ }\href
  {https://doi.org/10.1103/PhysRevB.61.5337} {\bibfield  {journal} {\bibinfo
  {journal} {Phys. Rev. B}\ }\textbf {\bibinfo {volume} {61}},\ \bibinfo
  {pages} {5337} (\bibinfo {year} {2000})}\BibitemShut {NoStop}%
\bibitem [{\citenamefont {Young}\ and\ \citenamefont
  {Rappe}(2012)}]{young2012}%
  \BibitemOpen
  \bibfield  {author} {\bibinfo {author} {\bibfnamefont {S.~M.}\ \bibnamefont
  {Young}}\ and\ \bibinfo {author} {\bibfnamefont {A.~M.}\ \bibnamefont
  {Rappe}},\ }\bibfield  {title} {\bibinfo {title} {First principles
  calculation of the shift current photovoltaic effect in ferroelectrics},\
  }\href {https://doi.org/10.1103/PhysRevLett.109.116601} {\bibfield  {journal}
  {\bibinfo  {journal} {Phys. Rev. Lett.}\ }\textbf {\bibinfo {volume} {109}},\
  \bibinfo {pages} {116601} (\bibinfo {year} {2012})}\BibitemShut {NoStop}%
\bibitem [{\citenamefont {Cook}\ \emph {et~al.}(2017)\citenamefont {Cook},
  \citenamefont {Fregoso}, \citenamefont {de~Juan}, \citenamefont {Coh},\ and\
  \citenamefont {Moore}}]{Cook_2017}%
  \BibitemOpen
  \bibfield  {author} {\bibinfo {author} {\bibfnamefont {A.~M.}\ \bibnamefont
  {Cook}}, \bibinfo {author} {\bibfnamefont {B.~M.}\ \bibnamefont {Fregoso}},
  \bibinfo {author} {\bibfnamefont {F.}~\bibnamefont {de~Juan}}, \bibinfo
  {author} {\bibfnamefont {S.}~\bibnamefont {Coh}},\ and\ \bibinfo {author}
  {\bibfnamefont {J.~E.}\ \bibnamefont {Moore}},\ }\bibfield  {title} {\bibinfo
  {title} {Design principles for shift current photovoltaics},\ }\href
  {https://doi.org/10.1038/ncomms14176} {\bibfield  {journal} {\bibinfo
  {journal} {Nat. Commun.}\ }\textbf {\bibinfo {volume} {8}},\ \bibinfo {pages}
  {14176} (\bibinfo {year} {2017})}\BibitemShut {NoStop}%
\bibitem [{not()}]{noteVortex}%
  \BibitemOpen
  \bibinfo {note} {These are vortices of the shift-vector field in momentum
  space; they should not be confused with the real-space vortices of the
  shift-current density discussed in Refs.~\cite{Hu_2023,
  Agarwal_2025}.}\BibitemShut {Stop}%
\bibitem [{\citenamefont {Cao}\ \emph {et~al.}(2018)\citenamefont {Cao},
  \citenamefont {Wu},\ and\ \citenamefont {Louie}}]{CaoWuLouie_2018}%
  \BibitemOpen
  \bibfield  {author} {\bibinfo {author} {\bibfnamefont {T.}~\bibnamefont
  {Cao}}, \bibinfo {author} {\bibfnamefont {M.}~\bibnamefont {Wu}},\ and\
  \bibinfo {author} {\bibfnamefont {S.~G.}\ \bibnamefont {Louie}},\ }\bibfield
  {title} {\bibinfo {title} {Unifying optical selection rules for excitons in
  two dimensions: Band topology and winding numbers},\ }\href
  {https://doi.org/10.1103/PhysRevLett.120.087402} {\bibfield  {journal}
  {\bibinfo  {journal} {Phys. Rev. Lett.}\ }\textbf {\bibinfo {volume} {120}},\
  \bibinfo {pages} {087402} (\bibinfo {year} {2018})}\BibitemShut {NoStop}%
\bibitem [{\citenamefont {Watanabe}\ and\ \citenamefont
  {Yanase}(2021)}]{Watanabe_2021}%
  \BibitemOpen
  \bibfield  {author} {\bibinfo {author} {\bibfnamefont {H.}~\bibnamefont
  {Watanabe}}\ and\ \bibinfo {author} {\bibfnamefont {Y.}~\bibnamefont
  {Yanase}},\ }\bibfield  {title} {\bibinfo {title} {Chiral photocurrent in
  parity-violating magnet and enhanced response in topological
  antiferromagnet},\ }\href {https://doi.org/10.1103/PhysRevX.11.011001}
  {\bibfield  {journal} {\bibinfo  {journal} {Phys. Rev. X}\ }\textbf {\bibinfo
  {volume} {11}},\ \bibinfo {pages} {011001} (\bibinfo {year}
  {2021})}\BibitemShut {NoStop}%
\bibitem [{\citenamefont {Wang}\ and\ \citenamefont
  {Braunstein}(2026)}]{WangBraunstein_2026}%
  \BibitemOpen
  \bibfield  {author} {\bibinfo {author} {\bibfnamefont {Z.-W.}\ \bibnamefont
  {Wang}}\ and\ \bibinfo {author} {\bibfnamefont {S.~L.}\ \bibnamefont
  {Braunstein}},\ }\href@noop {} {\bibinfo {title} {{Hodge} topology of
  semiclassical transport: A coordinate-free geometric framework for the
  anomalous {Hall} effect and non-linear {Berry} dipole}} (\bibinfo {year}
  {2026}),\ \Eprint {https://arxiv.org/abs/2606.31409} {arXiv:2606.31409
  [cond-mat.mes-hall]} \BibitemShut {NoStop}%
\bibitem [{\citenamefont {Fregoso}\ \emph {et~al.}(2017)\citenamefont
  {Fregoso}, \citenamefont {Morimoto},\ and\ \citenamefont
  {Moore}}]{Fregoso2017}%
  \BibitemOpen
  \bibfield  {author} {\bibinfo {author} {\bibfnamefont {B.~M.}\ \bibnamefont
  {Fregoso}}, \bibinfo {author} {\bibfnamefont {T.}~\bibnamefont {Morimoto}},\
  and\ \bibinfo {author} {\bibfnamefont {J.~E.}\ \bibnamefont {Moore}},\
  }\bibfield  {title} {\bibinfo {title} {Quantitative relationship between
  polarization differences and the zone-averaged shift photocurrent},\ }\href
  {https://doi.org/10.1103/PhysRevB.96.075421} {\bibfield  {journal} {\bibinfo
  {journal} {Phys. Rev. B}\ }\textbf {\bibinfo {volume} {96}},\ \bibinfo
  {pages} {075421} (\bibinfo {year} {2017})}\BibitemShut {NoStop}%
\bibitem [{\citenamefont {Haldane}(1988)}]{Haldane1988}%
  \BibitemOpen
  \bibfield  {author} {\bibinfo {author} {\bibfnamefont {F.~D.~M.}\
  \bibnamefont {Haldane}},\ }\bibfield  {title} {\bibinfo {title} {Model for a
  quantum {Hall} effect without {Landau} levels: Condensed-matter realization
  of the ``parity anomaly''},\ }\href
  {https://doi.org/10.1103/PhysRevLett.61.2015} {\bibfield  {journal} {\bibinfo
   {journal} {Phys. Rev. Lett.}\ }\textbf {\bibinfo {volume} {61}},\ \bibinfo
  {pages} {2015} (\bibinfo {year} {1988})}\BibitemShut {NoStop}%
\bibitem [{\citenamefont {McCann}\ and\ \citenamefont
  {Fal'ko}(2006)}]{McCann_2006}%
  \BibitemOpen
  \bibfield  {author} {\bibinfo {author} {\bibfnamefont {E.}~\bibnamefont
  {McCann}}\ and\ \bibinfo {author} {\bibfnamefont {V.~I.}\ \bibnamefont
  {Fal'ko}},\ }\bibfield  {title} {\bibinfo {title} {{Landau}-level degeneracy
  and quantum {Hall} effect in a graphite bilayer},\ }\href
  {https://doi.org/10.1103/PhysRevLett.96.086805} {\bibfield  {journal}
  {\bibinfo  {journal} {Phys. Rev. Lett.}\ }\textbf {\bibinfo {volume} {96}},\
  \bibinfo {pages} {086805} (\bibinfo {year} {2006})}\BibitemShut {NoStop}%
\bibitem [{\citenamefont {McCann}\ and\ \citenamefont
  {Koshino}(2013)}]{McCann_2013}%
  \BibitemOpen
  \bibfield  {author} {\bibinfo {author} {\bibfnamefont {E.}~\bibnamefont
  {McCann}}\ and\ \bibinfo {author} {\bibfnamefont {M.}~\bibnamefont
  {Koshino}},\ }\bibfield  {title} {\bibinfo {title} {The electronic properties
  of bilayer graphene},\ }\href {https://doi.org/10.1088/0034-4885/76/5/056503}
  {\bibfield  {journal} {\bibinfo  {journal} {Rep. Prog. Phys.}\ }\textbf
  {\bibinfo {volume} {76}},\ \bibinfo {pages} {056503} (\bibinfo {year}
  {2013})}\BibitemShut {NoStop}%
\bibitem [{\citenamefont {Fukui}\ \emph {et~al.}(2005)\citenamefont {Fukui},
  \citenamefont {Hatsugai},\ and\ \citenamefont {Suzuki}}]{Fukui_2005}%
  \BibitemOpen
  \bibfield  {author} {\bibinfo {author} {\bibfnamefont {T.}~\bibnamefont
  {Fukui}}, \bibinfo {author} {\bibfnamefont {Y.}~\bibnamefont {Hatsugai}},\
  and\ \bibinfo {author} {\bibfnamefont {H.}~\bibnamefont {Suzuki}},\
  }\bibfield  {title} {\bibinfo {title} {{Chern} numbers in discretized
  {Brillouin} zone: Efficient method of computing (spin) {Hall} conductances},\
  }\href {https://doi.org/10.1143/JPSJ.74.1674} {\bibfield  {journal} {\bibinfo
   {journal} {J. Phys. Soc. Jpn.}\ }\textbf {\bibinfo {volume} {74}},\ \bibinfo
  {pages} {1674} (\bibinfo {year} {2005})}\BibitemShut {NoStop}%
\bibitem [{\citenamefont {Zheng}\ \emph {et~al.}(2023)\citenamefont {Zheng},
  \citenamefont {Chang},\ and\ \citenamefont {Cheng}}]{Zheng_2023}%
  \BibitemOpen
  \bibfield  {author} {\bibinfo {author} {\bibfnamefont {Z.}~\bibnamefont
  {Zheng}}, \bibinfo {author} {\bibfnamefont {K.}~\bibnamefont {Chang}},\ and\
  \bibinfo {author} {\bibfnamefont {J.~L.}\ \bibnamefont {Cheng}},\ }\bibfield
  {title} {\bibinfo {title} {Gate voltage induced injection and shift currents
  in {AA}- and {AB}-stacked bilayer graphene},\ }\href
  {https://doi.org/10.1103/PhysRevB.108.235401} {\bibfield  {journal} {\bibinfo
   {journal} {Phys. Rev. B}\ }\textbf {\bibinfo {volume} {108}},\ \bibinfo
  {pages} {235401} (\bibinfo {year} {2023})}\BibitemShut {NoStop}%
\bibitem [{\citenamefont {Postlewaite}\ \emph {et~al.}(2024)\citenamefont
  {Postlewaite}, \citenamefont {Raj}, \citenamefont {Chaudhary},\ and\
  \citenamefont {Fiete}}]{Postlewaite_2024}%
  \BibitemOpen
  \bibfield  {author} {\bibinfo {author} {\bibfnamefont {A.}~\bibnamefont
  {Postlewaite}}, \bibinfo {author} {\bibfnamefont {A.}~\bibnamefont {Raj}},
  \bibinfo {author} {\bibfnamefont {S.}~\bibnamefont {Chaudhary}},\ and\
  \bibinfo {author} {\bibfnamefont {G.~A.}\ \bibnamefont {Fiete}},\ }\bibfield
  {title} {\bibinfo {title} {Quantum geometry and nonlinear optical responses
  in rhombohedral trilayer graphene},\ }\href
  {https://doi.org/10.1103/PhysRevB.110.245122} {\bibfield  {journal} {\bibinfo
   {journal} {Phys. Rev. B}\ }\textbf {\bibinfo {volume} {110}},\ \bibinfo
  {pages} {245122} (\bibinfo {year} {2024})}\BibitemShut {NoStop}%
\bibitem [{\citenamefont {Hu}\ \emph {et~al.}(2023)\citenamefont {Hu},
  \citenamefont {Naik}, \citenamefont {Chan}, \citenamefont {Ruan},\ and\
  \citenamefont {Louie}}]{Hu_2023}%
  \BibitemOpen
  \bibfield  {author} {\bibinfo {author} {\bibfnamefont {C.}~\bibnamefont
  {Hu}}, \bibinfo {author} {\bibfnamefont {M.~H.}\ \bibnamefont {Naik}},
  \bibinfo {author} {\bibfnamefont {Y.-H.}\ \bibnamefont {Chan}}, \bibinfo
  {author} {\bibfnamefont {J.}~\bibnamefont {Ruan}},\ and\ \bibinfo {author}
  {\bibfnamefont {S.~G.}\ \bibnamefont {Louie}},\ }\bibfield  {title} {\bibinfo
  {title} {Light-induced shift current vortex crystals in moir\'e
  heterobilayers},\ }\href {https://doi.org/10.1073/pnas.2314775120} {\bibfield
   {journal} {\bibinfo  {journal} {Proc. Natl. Acad. Sci. U.S.A.}\ }\textbf
  {\bibinfo {volume} {120}},\ \bibinfo {pages} {e2314775120} (\bibinfo {year}
  {2023})}\BibitemShut {NoStop}%
\bibitem [{\citenamefont {Agarwal}\ \emph {et~al.}(2025)\citenamefont
  {Agarwal}, \citenamefont {Jankowski}, \citenamefont {Bennett},\ and\
  \citenamefont {Slager}}]{Agarwal_2025}%
  \BibitemOpen
  \bibfield  {author} {\bibinfo {author} {\bibfnamefont {A.}~\bibnamefont
  {Agarwal}}, \bibinfo {author} {\bibfnamefont {W.~J.}\ \bibnamefont
  {Jankowski}}, \bibinfo {author} {\bibfnamefont {D.}~\bibnamefont {Bennett}},\
  and\ \bibinfo {author} {\bibfnamefont {R.-J.}\ \bibnamefont {Slager}},\
  }\bibfield  {title} {\bibinfo {title} {Shift photocurrent vortices from
  topological polarization textures},\ }\href
  {https://doi.org/10.1038/s41699-025-00552-6} {\bibfield  {journal} {\bibinfo
  {journal} {npj 2D Mater. Appl.}\ }\textbf {\bibinfo {volume} {9}},\ \bibinfo
  {pages} {43} (\bibinfo {year} {2025})},\ \Eprint
  {https://arxiv.org/abs/2408.04017} {arXiv:2408.04017 [cond-mat.mes-hall]}
  \BibitemShut {NoStop}%
\end{thebibliography}%

\end{document}